\documentclass[aps,prl,twocolumn,groupedaddress,showpacs]{revtex4}
\usepackage{amsmath}  
\usepackage{amsfonts} 
\usepackage{amssymb}
\usepackage{graphicx} 

 \usepackage{lstdoc,longtable}
 \usepackage{tabularx}

\begin{document}

\title{Defining of three-dimensional acceleration and inertial mass \\
leading to the simple form F=\textit{M}A of relativistic motion equation}

\author{Grzegorz M. Koczan\footnote{grzegorz\_koczan@sggw.edu.pl, gkoczan@fuw.edu.pl}}
\affiliation{Warsaw University of Life Sciences (WULS/SGGW), Poland }

  

\date{\today}

\begin{abstract}
Newton's II Law  of Dynamics is a law of motion but also a useful definition of force (F=\textit{M}A) or inertial mass (\textit{M}=F/A), assuming a definition of acceleration and parallelism of force and acceleration. In the Special Relativity, out of these three only the description of force (F=\textit{d}p/\textit{dt}) does not raise doubts. The greatest problems are posed by mass, which may be invariant rest mass or relativistic mass or even directional mass, like longitudinal mass. This results from breaking the assumption of the parallelism of force and standard acceleration. It turns out that these issues disappear if the relativistic acceleration A is defined by a relativistic velocity subtraction formula. This basic fact is obscured by some subtlety related to the calculation of the relativistic differential of velocity. It is based on the direction of force rather than on a transformation to an instantaneous resting system. The reference to a non-resting system generates a little different velocity subtraction formulae. This approach confirms Oziewicz binary and ternary relative velocities as well as the results of other researchers. Thus, the relativistic three-dimensional acceleration is neither rest acceleration, nor four-acceleration, nor standard acceleration. As a consequence, inertial mass in any direction of the force has the same value as relativistic mass. In other words, the concepts of transverse mass and longitudinal mass, which depend on velocity, have been unified.
In this work a full relativistic equation is derived for the motion of a body with variable mass whose form confirmed the previously introduced definitions. In addition, these definitions are in line with the general version of the principle of mass and energy equivalence. The work presents a detailed review and discussion of different approaches to the subject in relation to original historical and contemporary texts. On this basis, a proposal is made for a consistent definition of relativistic quantities associated with velocity change.

\


\end{abstract}








%




\maketitle 


\tableofcontents

\section{Introduction}
Special Relativity (SR) is well grounded, both theoretically and experimentally. However, within the dynamics of SR, there are some interpretation issues that are not fully agreed on by physicists. 
Disputes about these issues, although they sometimes sound explicitly or implicitly, are often ignored and reduced to philosophy, not to mathematics or experiment. Proponents of a particular convention and interpretation are convinced that it is the only right one. In addition, the same physicists claim that there is no collision in predicting the results of experiments under the alternative conventions. If these conventions are to be physically equivalent, then how is one of them correct and the other not? This logical inconsistency shows that in SR there are issues that have not been resolved until now.

   The most famous problematic issue is the topic of relativistic (total) mass versus rest mass ($M$ vs $m$). The physicists of elementary particles try to treat mass as a constant parameter describing a given particle. At the same time, they display carelessness in relation to the inertial mass in the mechanics. Einstein's demand for a good definition of mass (depending on velocity) has not been implemented and is misinterpreted. This problem affects the interpretation and even the mathematical form of the famous principle of mass and energy equivalence ($E_0=mc^2$ vs $E=Mc^2$ or $E=\gamma mc^2$). Thus, in analogy to the weak and strong principle of equivalence of inertial and gravitational mass, we can speak of weak-resting ($E_0=mc^2$) and strong-general ($E=Mc^2$) version of the principle of mass and energy equivalence. The strong-general version, which takes into account the kinetic and potential energy (binding energy) is not physically equivalent to the formula $E=\gamma mc^2$ or the formula $E=\sqrt{(mc^2)^2+(\mathbf{p}c)^2}$. Of course, binding energy can also be included in $E_0$, but this does not change the fact that $E$ is more general. Since the principle of energy conservation applies, all forms of energy should have an equal contribution to the mass of the system.
	
    A somewhat less spectacular but closely related issue is the dispute over the most general formula of the second law of dynamics. It is claimed that the simple product form of this law known from the school, i.e. $\mathbf{F}=m\mathbf{a}$ (or $\mathbf{F}=M\mathbf{A}$), is less general than the form with derivative of momentum with respect to time (rate form $\mathbf{F}=d\mathbf{p}/dt$). Is this really true? It turns out that in a sense no, and this negative answer enables for an unambiguous definition of the inertial mass. To justify a negative answer to this question, it is necessary to analyze the variable mass in the Newton and SR mechanics. In the case of SR, the difficulty is at least doubled, because a variable rest mass $m(t)$ should be considered, which further complicates the variability of the relativistic mass or other mass $M(t,\mathbf{v})$. Furthermore, it is not clear what the simple product form of the ordinary motion equation in SR is ($\mathbf{F} \neq m\mathbf{a}$ and $M=?$ , $\mathbf{A}=?$). What is more, the full general relativistic equation of motion with variable rest mass is not known yet (for $\mathbf{u} \not\,\parallel \mathbf{v}$ or $\mathbf{F}_{ext} \neq 0$). 
				
   Another problem of SR is the three-dimensional or four-dimensional duality of physical vector quantities (e.g. $\mathbf{v}$, $\mathbf{F}$ versus $\nu^{\mu}$, $f^{\mu}$). This dualism boils down to the equality of the spatial part of the four-vector and the three-vector practically only for position and momentum. However, already for velocity or force, spatial parts of the dual quantities differ by the Lorentz factor ($\vec{\nu}=\gamma\mathbf{v}$, $\vec{f}=\gamma\mathbf{F}$). Formulating SR in Minkowski four-dimensional spacetime using four-vector is a widely recognized, elegant geometric method. However, one cannot forget that SR is a physical theory (it is part of physics) and should refer as closely as possible to observable physical reality. Unfortunately, the nature of this physical reality is at least seemingly three-dimensional, not directly four-dimensional. The fourth dimension, which is time, is not perceived as a spatial dimension, and SR does not explain its temporary limitation to a specific local value called the present. SR also does not have the time arrow highlight. In view of the above, relativistic physics can, and even should, be formulated in three-dimensional terms. This state of affairs is confirmed by quantum mechanics. It turns out that relativistic quantum mechanics forces the choice of three-dimensional space in a significant distinction from the time coordinate. This is evidenced by the lack of a time operator in virtually all quantum mechanics, while three-dimensional position operators exist. The situation is even more vivid in quantum gravity. Well, the Wheeler--DeWitt equation describing quantum spacetime does not contain the fourth dimension at all, the notion of time. Modern specialists in quantum loop gravity are trying to derive the concept of time from the entanglement of three-dimensional quantum space. An additional argument in favor of three-vectors versus four-vectors is that both types of vectors contain essentially three independent parameters. The only exception is the trivial four-position vector. However, the four-vectors of velocity and momentum are normalized, and the four-vectors of acceleration and force are orthogonal to velocity. These conditions reduce the number of degrees of freedom from four to three.
   
In the case of ordinary acceleration, the mentioned correspondence of dual quantities is more complicated than for other vectors. This correspondence can be greatly simplified by the definition of three-dimensional relativistic acceleration $\mathbf{A}$. The definition of this acceleration is not an {\it ad hoc} definition based only on the relation of force and relativistic mass ($\mathbf{A}=\mathbf{F}/M$). The notion of acceleration is based on the velocity differential $(d\mathbf{v})_{rel}$, which should actually be calculated not as an ordinary difference, but in accordance with the velocity algebra in SR kinematics. The essence of Einstein's relativistic velocity subtraction (operation $\ominus_0$) is the Lorentz transformation into a system moving at a subtracted velocity (means into instantaneous rest reference system for velocity differential). However, if for some reasons, the transformation will be performed to slightly different reference systems (than resting), the subtraction of velocity will take on another form (e.g. differential-equivalent operations $\ominus_{\parallel}$ or $\ominus_{\perp}$). When calculating the relativistic velocity differential in the context of acceleration and motion equation, the Lorentz transformation generally does not refer only to the instantaneous resting reference system. The correct reference system (according to $\ominus_{\parallel}$) depends on the direction in which the differential of velocity  is calculated. For example, in the direction perpendicular to the velocity, no Lorentz transformation is needed at all, while in the direction parallel to the velocity, we need transformation to the instantaneous resting system of the body. In the direction of force, however, the Lorentz transformation is performed at a velocity projected on the direction of force. Furthermore, the interpretation of operation $\ominus_{\wedge}$ in calculating the velocity differential is based on the Lorentz transformation with an infinitely low velocity (differential), so it is asymptotically associated with the initial reference system, and not with rest system of the body. This method (operation $\ominus_{\wedge}$) of subtracting velocity leads to the most general concept of Oziewicz ternary relative velocity. And the operation $\ominus_{\perp}$ resulting from calculating the velocity of jet gases relative to the rocket leads to Oziewicz binary velocity. All differential-equivalent operations ($\ominus_{\parallel}$, $\ominus_{\perp}$, $\ominus_{\wedge}$, $\ominus_{\vee}$, $\boxminus$) are also equivalent for acceleration. The existence of many  differential-equivalent operations can be explained by the fact that the relativistic velocity differential is not an exact differential. Consequently, finite velocity differences depend on the contractual integration path. This situation is reminiscent of the property of noncommutability in the Lorentz group and is associated with the Thomas-Wigner rotation.
It turns out that the subtleties of velocity vector compositions are described by mathematical physicists in terms of relativistic groupoid or a loop (quasigroup) called a gyrogroup. These structures are based on hyperbolic geometry applied to the concept of relativistic relative velocity.
	
Therefore, this work does not introduce alternative content to SR. On the contrary, it deepens the conceptual apparatus and solves the above-mentioned problems in an orthodox, but original way.  It has the following division of contents: sections I--V constitute a strictly historical introduction, sections VI--VIII present the {\it status quo} of the subject matter and sections IX--XIV present the developments made by the author. The historical part contains original and precise analysis of pioneering research, e.g. the Kaufmann mass measurement experiment or various Einstein mass definitions. The {\it status quo} presentation contains a description of the criticism and polemical disputes. The author's results are confronted with contemporary results.

\section{I. Nonrelativistic motion equation} 

In 1687, Newton published the second law of motion equivalent to the equation \cite{Newton}:
\begin{equation}
\label{Newton rate}
\mathbf{F}=\frac{d\mathbf{p}}{dt}=\frac{d(m\mathbf{v})}{dt},
\end{equation}
where: $\mathbf{F}$ -- force, $\mathbf{p}$ -- momentum, $t$ -- time, $m$ -- mass (invariant), $\mathbf{v}$ -- velocity. For constant mass Newton equation adopts the form well-known from school: 
\begin{equation}
\label{Newton simple}
\mathbf{F}=m \frac{d\mathbf{v}}{dt}=m\mathbf{a},
\end{equation}
where: $\mathbf{a}$ -- acceleration. The equation (\ref{Newton rate}) will here be referred to as the rate form and the equation (\ref{Newton simple}) will be referred to as the simple form. This last form was used by Mach in 1883 in his system of mechanical definitions: \textit{Moving force is the product of the mass-value of a body into the acceleration inducted in that body} \cite{Mach}.
  
A naive application of the rate form for a body with variable mass leads to a wrong equation:
\begin{equation}
\label{incorrect}
\mathbf{F}=\frac{d(m\mathbf{v})}{dt}=m\frac{d\mathbf{v}}{dt}+\frac{dm}{dt}\mathbf{v}\:\:\:\text{(incorrect)}. 
\end{equation}
This equation is inconsistent with the Galilean transformation \cite{Pesce}. The correct equation for the movement of a body with variable mass was not discovered until the 19th century:
\begin{equation}
\label{rocket rate}
\mathbf{F}_{ext}=\frac{d\mathbf{p}_{sys}}{dt}=\frac{d(m\mathbf{v}+\delta m \mathbf{u})}{dt}=m\frac{d\mathbf{v}}{dt}-\frac{dm}{dt}(\mathbf{u}-\mathbf{v}),
\end{equation}
where: $\mathbf{F}_{ext}$ -- external force, $\mathbf{p}_{sys}$ -- momentum of system, $\delta m$ -- small attaching (or detaching) mass, $\mathbf{u}$ -- velocity of $\delta m$.
The earliest known publication with the formula (\ref{rocket rate}) is the work from 1812 by von Buquoy \cite{Buquoy 1812}, continued in 1815 \cite{Buquoy 1815}. The problem of variable mass was also studied by a well-known scientist Poisson \cite{Poisson} in 1819. However, the equation (\ref{rocket rate}) or (\ref{rocket simple}) is best known as Meshchersky equation from his publication of 1897 \cite{Mieszczerski 1897} or 1904 \cite{Mieszczerski 1904}. On the basis of this equation Tsiolkovsky derived the famous rocket equation published in 1903 \cite{Ciolkowski}.

The second law of variable mass motion in the quasirate form (\ref{rocket rate}) can be put in a simple form:
\begin{equation}
\label{rocket simple}
m \mathbf{a}=\mathbf{F}_{tot}=\mathbf{F}_{ext}+\mathbf{F}_{th}=\mathbf{F}_{ext}+\frac{dm}{dt}(\mathbf{u}-\mathbf{v}),
\end{equation} 
where: $\mathbf{F}_{tot}$ -- total force, $\mathbf{F}_{th}$ -- thrust force (or braking).
This also applies to the thrust force expression if we use the third law of dynamics:
\begin{equation}
\label{thrust}
\mathbf{F}_{th}=-\mathbf{F}_{\delta m}=-\lim_{\Delta t\rightarrow0} \delta m \frac{\mathbf{u} - \mathbf{v}}{\Delta t}, 
\end{equation}
where: $\mathbf{F}_{\delta m}$ -- force acting on $\delta m=-\Delta m$, $\Delta m$ -- mass increment (implicitly negative), $\Delta t$ -- time increment. According to Johnson \cite{Johnson}, the problem of variable mass was also solved by Moore in 1813 \cite{Moore}. Let's try to write the equation in simple form (\ref{rocket simple}) back in the rate form:
\begin{equation}
\label{rocket rate 1}
\frac{d(m\mathbf{v})}{dt}=\mathbf{F}_{tot}+\frac{dm}{dt}\mathbf{v}=\mathbf{F}_{ext}+\frac{dm}{dt}\mathbf{u}.
\end{equation} 
Therefore, the derivative of the momentum of the rocket is neither total force nor external force, and none of the additions to these forces is here a thrust force.

The presented considerations show that in nonrelativistic mechanics the rate form equation is not more general than the simple form (see also \cite{Wolny}). On the contrary, (\ref{incorrect}) is false, and the equations (\ref{rocket simple}) and (\ref{thrust}) in the simple form are true. And the quasirate form (\ref{rocket rate}) describes a closed system with constant mass, not solely a body with variable mass. So equation (\ref{rocket rate 1}) can only be called a pseudo-rate form.

\section{II. Lorentz investigation 1899, 1904}

As a result of analyses of different reference systems for the movement of a charge in an electromagnetic field, in 1899 Lorentz described the dependence of inertial mass on velocity and direction \cite{Lorentz 1899}:
\begin{equation}
\label{Lorentz masses}
\mu_{\parallel}=\frac{\mathrm{F}_{\parallel}}{\mathrm{a}_{\parallel}}=\gamma^{3}m\:, \: \mu_{\perp}=\frac{\mathrm{F}_{\perp}}{\mathrm{a}_{\perp}}=\gamma m,
\end{equation} 
where: $\mu_{\parallel}$ -- longitudinal mass, $\mu_{\perp}$ -- transverse mass, $\gamma=1/(1-\mathrm{v}^2/c^2)^{1/2}$ -- Lorentz factor, $c$ -- speed of light, $\mathrm{F}_{\parallel}$ -- component of vector parallel to velocity (for force), $\mathrm{a}_{\perp}$ -- component of vector perpendicular to velocity (for acceleration). These formulas were presented by Lorentz in a text at the end of his work from 1899 and then included in his work from 1904 \cite{Lorentz 1904} in a form that looks like a vector (of mass). In both works Lorentz also considered additional scaling factor of spacetime coordinates (dilatation), which in this work equal 1. Surprisingly, correct formulas (\ref{Lorentz masses}) were obtained by Lorentz on the basis of the not perfectly explicit and clear transformation \cite{Lorentz 1904, Lorentz 1899}: 
\begin{equation}
x'=\gamma \tilde{x} \:,\:t'=\frac{1}{\gamma}t - \gamma \frac{\mathrm{v}}{c^{2}}\,\tilde{x} \:\:\:\:\: \text{(unclear)},
\end{equation} 
where: $x',\ t'$ -- position and time in a moving system; $x,\ t$ -- position and time in an initial system, and $\tilde{x}\equiv x-\mathrm{v} t$ is implicitly the position in a moving system calculated according to the Galileo transformation (originally denoted by $x$) \cite{Damour}.
Interestingly, the original transformation denoted by Lorentz ($x$ without $\tilde{x}$) with accuracy of the following inversion $x\rightarrow - ct'$, $x'\rightarrow - ct$, $t\rightarrow x'/c$, $t'\rightarrow x/c$ is in accordance with transformation given by Tangherlini in 1958 \cite{Tangherlini}, Mansouri and Sexl in 1977 \cite{Mansouri} and some other contemporary independent scholars \cite{reference system}. It turns out that such a modified transformation can be explained within SR by means of appropriate clocks synchronization, in this case external synchronization \cite{Lammerzahl}. 
The Lorentz transformation was simplified (implicitly equivalent) and clearly recorded on 5th June 1905 by Poincare \cite{Poincare}, who gave it a form very close to its contemporary version:
\begin{equation}
\label{Lorentz Poincare}
x'=\gamma (x-\mathrm{v}t) \:,\:t'=\gamma \left(t - \frac{\mathrm{v}}{c^{2}}\,x \right) \:,\: y'=y \:,\: z'=z,
\end{equation}
where: $y, z, y', z'$ -- coordinates along the directions perpendicular to velocity.
Poincare originally used units like $c=1$, and additionally, just like Lorentz, he considered scaling (dilatation), which is omitted here.

\section{III. Kaufmann and Abraham 1901, 1902}

In 1897 Searle \cite{Searle} calculated the energy ${E}$ (originally denoted by $\mathrm{W}$) of electromagnetic field generated by a sphere or an ellipsoid in uniform movement. This energy was not proportional to velocity squared, which suggested the existence of rest energy and so-called electromagnetic mass which depends on velocity. The simplest way to calculate mass from energy was the formula used by Kaufmann in 1901 \cite{Kaufmann 1901}:

\begin{equation}
\label{Kaufmann}
\mu=\frac{1}{\mathrm{v}}\frac{dE}{d\mathrm{v}}\:\:\:\:\left(=\mu_{\parallel}=\frac{1}{\mathrm{v}}\frac{dE_k}{d\mathrm{v}}\right)\:,
\end{equation}
where: $\mu$ -- electron mass (longitudinal, electromagnetic), $E$ -- electron electromagnetic field energy (or relativistic energy), $E_k$ -- kinetic energy. 

It was only later when it turned out that the said formula determined the so-called longitudinal mass ($\mu=\mu_{\parallel}$). The terms of longitudinal mass and transverse mass were introduced using momentum by Abraham in 1902 \cite{Abraham 1}:
\begin{equation}
\label{Abraham}
\mu_{\parallel}=\frac{d\mathrm{p}}{d\mathrm{v}} \:\:, \: \:\mu_{\perp}=\frac{\mathrm{p}}{\mathrm{v}},
\end{equation} 
where: $\mathrm{p}$ -- electron electromagnetic field momentum (or electron momentum).
He also gave other formulae based on the Lagrangian function $L$ of the electron field (or electron itself):
\begin{equation}
\label{Abraham two}
\mu_{\parallel}=\frac{d^2L}{d\mathrm{v}^2} \:\: , \: \: \mu_{\perp}=\frac{1}{\mathrm{v}}\frac{dL}{d\mathrm{v}},
\end{equation} 
which was recognised in the next work by Abraham \cite{Abraham 2} from the same year. It should be noted that Abraham apart from Lagrangian in (\ref{Abraham two}) distinguished in (\ref{Kaufmann}) electrostatic energy (resting energy) and magnetic energy (simplified kinetic energy of motion). It was somewhat related to the longitudinal and transverse mass, but also generated an additional division into electromagnetic mass and specific mass (see e.g \cite{Morozov}). However, as part of the dynamics of the SR, this division does not seem necessary here. Similarly, for $E$, $p$, $L$ no distinction is made between field and particles designations, although they may differ and lead to different formulae for masses.

Kaufmann mass calculated from Searl's energy and Abraham's masses, which were calculated for the spherical electron, satisfies the inequality:
\begin{equation}
\mu_{\parallel}\leq\frac{m}{\left(1-\frac{\mathrm{v}^2}{c^2}\right)^{\frac{6}{5}}}\:\:, \:\: \mu_{\perp}\leq\frac{m}{\left(1-\frac{\mathrm{v}^2}{c^2}\right)^{\frac{2}{5}}}\:\: (\text{incorrect}).
\end{equation} 
The original equalities were more complex and contained logarithms. The simpler inequities given here are correct in relation to the equalities of Abraham, but they contradict the correct formulae of Lorentz (\ref{Lorentz masses}). The source of this contradiction was the omission of the longitudinal contraction of the spherical electron in motion. Abraham derived formulae with equalities for sphere in its first publication \cite{Abraham 1} and finally concluded them in his third work from 1903 \cite{Abraham 3}.

In 1901 Kaufmann \cite{Kaufmann 1901} was experimenting with the dependence of mass on velocity for radium beta radiation (see {\bf Tab. \ref{Kaufmann data}}). 
\begin{table}[h!]
\centering
\caption{Kaufmann's measurements results for calculation of dynamic (transverse) mass. Characteristic parameters and measured values of the experiment refer directly to the work \cite{Kaufmann 1901}. However, the velocity and the ratio of dynamic mass to charge were calculated all over again (for two variants of assumptions). Kaufmann originally gave the reverse charge-to-mass ratio, but similarly to the velocity calculations, he reproduced a fatal error. This error was corrected in another work \cite{Kaufmann 1902} containing unfortunately another unjustified modification.}
\begin{ruledtabular}

\begin{tabular}{c @{\hspace{1.4\tabcolsep}} c @{\hspace{1.4\tabcolsep}} c @{\hspace{1.4\tabcolsep}} c @{\hspace{1.4\tabcolsep}} c @{\hspace{1.4\tabcolsep}} c}
\multicolumn{2}{c}{Original results } & \multicolumn{2}{c}{Compiled results} & \multicolumn{2}{c}{ Recomp.  results} \\
\multicolumn{2}{c}{Kaufmann 1901} & \multicolumn{2}{c}{Koczan  2019} &  \multicolumn{2}{c}{$h'=$  1.873  $ cm $}  \\
\hline	
$z_0$ & $y_0$  & $\mathrm{v}$  & $M/e$  & $\mathrm{v}'$  & $(M/e)'$  \\
 $[cm]$ &  $[cm]$ & $[10^8 \frac{m}{s}]$ &  $[10^{-12} \frac{kg}{C}]$ &  $[10^8 \frac{m}{s}]$ & $[10^{-12} \frac{kg}{C}]$ \\
\hline
0.271	& 0.0621	& 2.82	& 16.14	& 2.97	& 15.29 \\
0.348	& 0.0839	& 2.68	& 13.32	& 2.82	& 12.63 \\
0.461	& 0.1175	& 2.53	& 10.79	& 2.67	& 10.23 \\
0.576	& 0.1565	& 2.37	& 9.39	& 2.50	& 8.89 \\
0.688	& 0.198	& 2.24	& 8.51	& 2.36	& 8.06 \\
\hline
\hline
$x_1$ & $x_2$ & $\delta$ & $U$ & $\mathrm{B}$ & $h$ \\
$[cm]$ & $[cm]$ & $[cm]$ & $[V]$ & $[T]$ & $[cm]$\\
\hline
2.07 & 2  & 0.1525 & 6750 & 0.0299 & 1.775 \\
\end{tabular}
\end{ruledtabular}
\label{Kaufmann data}
\end{table}
\begin{figure}[h!]
\centering
\includegraphics[width=9cm]{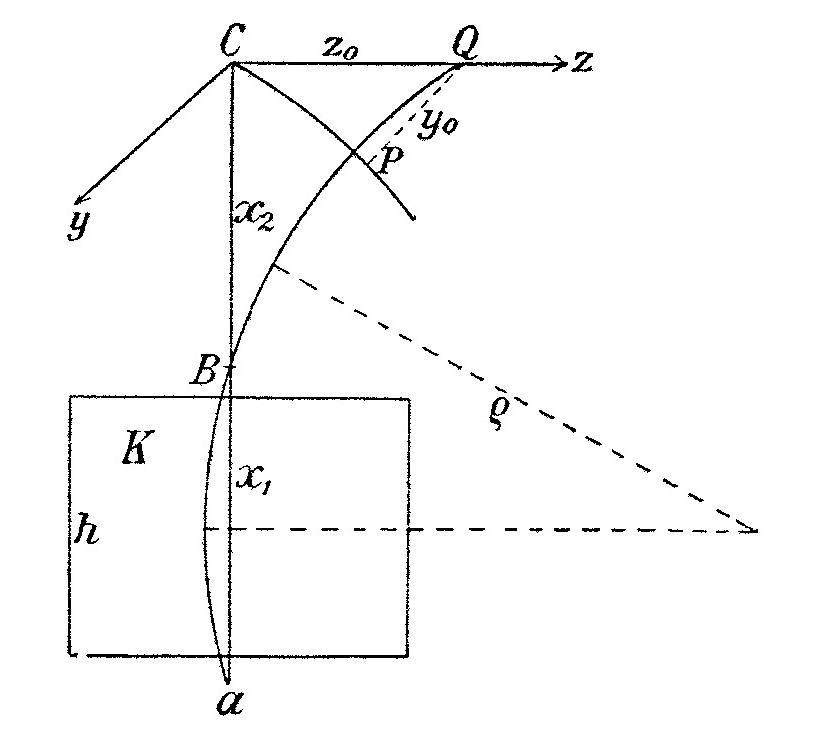}
\caption{A print from Kaufmann's work from 1901 \cite{Kaufmann 1901} showing schematically the electron trajectory in the experiment. The homogeneous magnetic field was directed in the opposite direction to the $y$ axis and covered the entire area of electron movement. The homogeneous electric field was parallel to the magnetic one but only covered the $K$ capacitor area. The electron that was recorded at the point $P(0, y_0, z_0)$ on the photographic plate $yz$ had to pass through the hole $B$, flying out earlier from the radioactive source located below the capacitor. The capacitor was thin enough ($\delta/x_1\approx0.07$) so that it can be roughly considered that the electrostatic force was perpendicular to the velocity of the electron. In addition, the electron exited the capacitor at exactly the same speed at which it fell into it. Thus, the experiment determined the transverse mass, which was approximately constant during the electron movement.}
\label{diagram 0}
\end{figure}

He measured specific charge $e/M$ of fast electrons on the basis of trajectory deviation in parallel electric and magnetic fields (see \textbf{Fig. \ref{diagram 0}}). The symbol $M$ refers here to apparent dynamic (transverse) mass ($\mathrm{F}/\mathrm{a}$ for $\mathbf{F}\perp\mathbf{v}$).  Electric field of the value $\mathrm{E}=U/\delta$ acted on the electron for about half of its trajectory. Next, the electron was deviated only by the magnetic field of the value $\mathrm{B}$ and registered on a photographic plate. The main measuring device was slightly over 4 cm long. An analysis of the photographic plate required the use of a microscope micrometer. 
Unfortunately, in 1901 Kaufmann miscalculated the radius $\rho$ of curvature of electrons trajectory. He corrected the mistakes in his next work in 1902 \cite{Kaufmann 1902}, in which he additionally rescaled the data, which brought his results closer to Abraham theory. The intervention in the data was considerable enough to lead Kaufmann to remove one of the five experimental points (the one with the highest velocity). The results compiled in {\bf Tab. \ref{Kaufmann data}} were obtained in a very similar method as used by Kaufmann in 1901. Basically, only the error in counting the radius of curvature has been corrected and greater precision of calculations has been applied. Apart from radius correction, the modified Kaufmann's calculation method from 1902 was not used.
 \begin{figure}[h!]
\centering
\includegraphics[width=9cm] {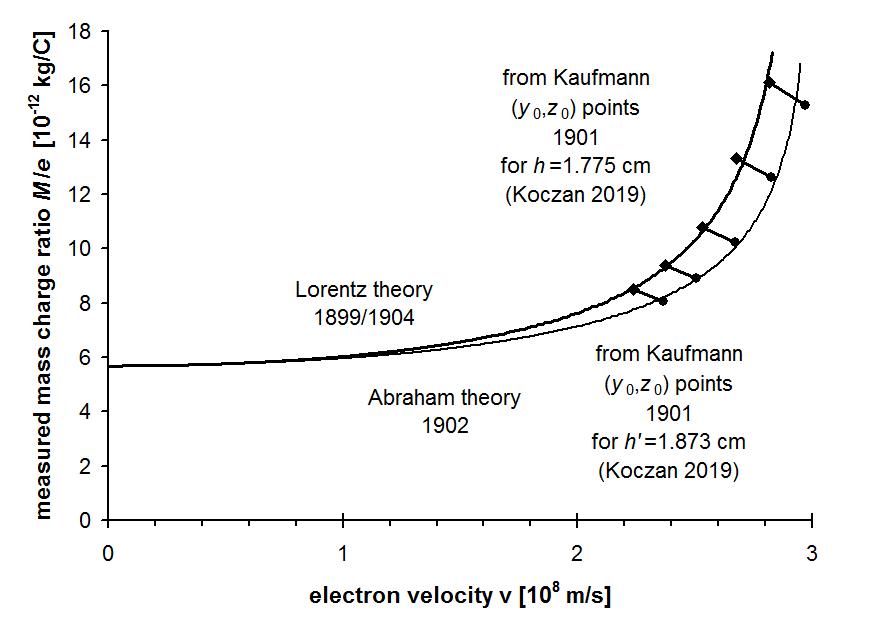}
\caption{The dependence of ratio of dynamic mass (transverse) and electron charge on its velocity on the basis of the first measurements by Kaufmann from 1901 \cite{Kaufmann 1901}. Five experimental points are blurred into sections depending on how far the electric field goes beyond the plates of the capacitor. The calculation of the left ends of the sections neglects the electric field outside the capacitor, and the right ones assume its bilateral exit by 1/3 of the distance between the plates and the aperture with the hole for electrons. Correct calculations did not require the use of relativistic mechanics and knowledge of physical constants.}
\label{diagram 1}
\end{figure}

Between 1902 and 1906 Kaufman continued his research, but because of the lack of determining an unambiguous value of magnetic field induction (among others) he did not solve the problem \cite{Kaufmann 1902 X, Kaufmann 1903, Kaufmann 1905, Kaufmann 1906}. However, Kaufmann significantly improved the precision of experimental points by using a strong source in the form of pure radium chloride $\mathrm{RaCl}_2$. He received this source from Maria Sk\l{}odowska-Curie and Pierre Curie \cite{Kaufmann 1902 X}.

The measurements results by Kaufmann were analysed in 1904 by Lorentz \cite{Lorentz 1904}. However, his evaluation was not exhaustive and unambiguous, for as he wrote: \textit{I have not found time for calculating the other tables in Kaufmann's paper}. Lorentz analysed tables III and IV from \cite{Kaufmann 1902 X} and tables II and III from \cite{Kaufmann 1903}. At the end of his analysis, Lorentz wrote: \textit{We may expect a satisfactory agreement with my formulae}. In 1906 Kaufmann's works \cite{Kaufmann 1906} (table VI, VII) were analysed by Planck \cite{Planck mass}. Despite the line of thinking drawn with his previous article \cite{Planck}, Planck had to admit a slight advantage of Abraham theory: \textit{It can be seen that the latter are closer to the sphere theory than to the relative theory}. Still, because of insufficient precision of the measurements, Planck did not consider the Kaufmann’s conclusion of his critic of Lorentz and Abraham theories as final. 

A thorough review of Kaufmann's works and their analysis by Lorentz and Planck (and even Einstein) was made by Cushing in 1981 \cite{Cushing}. Cushing's work contains many important drawings and tables regarding the Kaufmann experiments. Rather, this work confirms after Lorentz and Planck the lack of conclusiveness of the Kaufmann experiments regarding Abraham's theory in relation to Lorentz's theory. However, Cushing's oryginal analysis of the Kaufmann experiment from 1901 was based on too many approximations.

The subtlety of distinguishing Lorentz theory from Abraham theory is best evident from the diagram on {\bf Fig. \ref{diagram 1}} drawn on the basis of the first ever Kaufmann table \cite{Kaufmann 1901}. Contrary to the analyzes by Lorentz and Planck, the diagram takes into account the calculations of the ratio of mass (transverse) and charge from the experiment. Also, the points on the graph do not use any physical constants values $m, e$ and $c$ and all given data are calculated from experiment. However, if we assume the value of $c=3\cdot 10^8m/s$ to calculate the proper electron charge $e/m=\sqrt{1-\mathrm{v}^2/c^2}\ e/M$ based on {\bf Tab. \ref{Kaufmann data}} then we get:
\begin{equation}
e/m=(1.738\pm 0.021)\cdot 10^{11} C/kg \:\:\:\:\: \text{(reanalysis)},
\end{equation}
while the accepted value is $1.75888\cdot 10^{11} C/kg$.
Therefore, Kaufmann’s raw data fit better with Lorentz theory. However, this conformity fails if we assume that electric field is slightly outside the capacitor area ($h<h'<x_1$). Based on Cushing \cite{Cushing}, Kaufmann measured this, and Planck approximated the field decay at the edge of the capacitor with a linear function. The averaging of the Planck's approximation is equivalent to $h'\approx1.07h$, and here $h'\approx1.06h$ is assumed. However, due to the compliance with the Lorentz theory of the $h$ version, and not $h'$, the correctness of the version with the raw data $h$ can be considered post factum. Thus, it can be said that the reanalysis of the first Kaufmann experiment of 1901 showed post factum compliance with Lorentz's theory and not with Abraham's.

Officially, the experimental confirmation of the superiority of Lorentz theory over Abraham theory came with the experiments by Butcherer in 1908 \cite{Bucherer}, who used perpendicular, not parallel, magnetic and electric fields \cite{Wroblewski}. It is worth adding that Bucherer in 1904 \cite{Bucherer 1904}, as well as Langevin in 1905 \cite{Langevin} gave yet another model of electromagnetic mass, which one was not confirmed. In the Bucherer--Langevin model, the transverse mass was $\gamma^{2/3}m$. In practice, their formula did not differ much from the Abraham formula, and for $\mathrm{v}=0.9650c$ both formulae will be equal. While the Abraham model was based on an invariably spherical electron, the Bucherer--Langevin model assumed sphere deformation during motion that retained volume \cite{Janssen}. However, according to SR, the sphere should flatten in the direction of movement, and its volume should then decrease. The smaller size of the charge is the greater energy of the electromagnetic field around it, which may explain why the theory with the largest (transverse) mass has been experimentally confirmed. Contrary to appearances, Lorentz did not explicitly derive his theory from full calculations of energy of the flattened sphere, but from transformational rules (somewhat like Einstein later).

\section{IV. Einstein and Planck 1905--1907}

On 30 June 1905, Albert Einstein published his work \cite{Einstein June} where, based on the postulate of constant speed of light, he derived the correct transformation of coordinates in the form of (\ref{Lorentz Poincare}). Based on that he determined the transformation of the component of velocity parallel to the motion of frame (relativistic velocity subtraction formula): 
\begin{equation}
\label{velocity 0}
 (\mathrm{u}-\mathrm{v})_{rel}=\mathrm{u}\ominus\mathrm{v}=\mathrm{u'}=\frac{\mathrm{u}-\mathrm{v}}{1-\mathrm{u}\mathrm{v}/c^{2}},
\end{equation}
where: $\mathrm{u}$ -- body velocity, $\mathrm{v}$ -- boost velocity, $\mathrm{u}'$ -- body velocity in a new frame. Originally, Einstein wrote the law for addition, not subtraction of velocity. He also proposed a general expression for the value of resultant velocity vector $|\mathbf{u}\ominus_0 \mathbf{v}|$. For further applications, consider here the (seemingly) equivalent vector form of this velocity subtraction (see \cite{Mexico} for $\mathbf{u}\ominus_0\mathbf{v}:=(-\mathbf{v})\oplus\mathbf{u}$): 
\begin{equation}
\label{velocity}
\mathbf{u}\ominus_0\mathbf{v}=\mathbf{u'}=\frac{\mathbf{u}-\mathbf{v}+\frac{\gamma_{\mathrm{v}}}{\gamma_{\mathrm{v}}+1}(\mathbf{u}\times\mathbf{v})\times\mathbf{v}/c^{2}}{1-\mathbf{u}\mathbf{v}/c^{2}}=:\mathbf{u}_0,
\end{equation}
where Lorentz factor $\gamma_{\mathrm{v}}$ depends on $\mathrm{v}$.
This formula is called Einstein's relativistic law of velocity composition (subtraction or addition). In fact, Einstein, specifying this law only for the velocity value, determined it with the accuracy of rotation. Thus, in a sense, Einstein left a gap here for Thomas-Wigner rotation.
Formula (\ref{velocity}) is a vector generalization of the formula (\ref{velocity 0}), but $(\mathbf{u}\ominus_0\mathbf{v})_i \not\equiv \mathrm{u}_i \ominus \mathrm{v}_i$. Therefore, index 0 was used for the vector operation. Zero means that this is not the only possible way to approach the problem of subtraction of velocity vectors. The simplest, but not entirely correct vector generalization (depending on the choice of the Cartesian $xyz$ system) is the transfer of operation (\ref{velocity 0}) to the components as $(\mathbf{u}\ominus_1\mathbf{v})_i := \mathrm{u}_i \ominus \mathrm{v}_i$. 

Another important element of the work \cite{Einstein June} was the derivation of relativistic equation of motion for a charge in electromagnetic field: 
\begin{equation}
\label{Einstein dynamics}
m \gamma^{3} \mathbf{a}_{\parallel}=q\mathbf{E}_{\parallel} \:,\: m \gamma^{2}\mathbf{a}_{\perp}=\gamma q (\mathbf{E}_{\perp}+\mathbf{v}\times\mathbf{B}),
\end{equation}
where: $\mathbf{E}$ -- intensity of electric field, $\mathbf{B}$ -- induction of magnetic field, $q$ -- electric charge.
The method for derivation those equations, as well as their form, referred to the components of force vector in a resting frame of a body. So Einstein explicitly determined longitudinal and transverse mass as follows:
\begin{equation}
\label{Einstein masses}
\mu_{\parallel}=\frac{\mathrm{F'}_{\parallel}}{\mathrm{a}_{\parallel}}=\gamma^{3}m\ , \ \ \mu'_{\perp}=\frac{\mathrm{F'}_{\perp}}{\mathrm{a}_{\perp}}=\gamma^{2} m \ \ \ (\text{unclear}),
\end{equation}
where: $\mathrm{F'}_{\parallel}, \mathrm{F'}_{\perp}$ -- values of force components in a rest frame, $\mu'_{\perp}$ -- Einstein's transverse mass. 
Wanting to reduce the difference between longitudinal and transverse mass, Einstein choose the definition overstating the transverse mass by gamma factor. He was referring there to the rest value of force, at the same time using not rest acceleration (see \cite{Bazanski}). The factor $\gamma^2$ does occur with mass in the perpendicular component of four-force and in alternative energy based on four-force instead of force \cite{Osiak}, but Lorentz proposal (\ref{Lorentz masses}) was more correct. If Einstein had used in (\ref{Einstein masses}) primes in the denominator instead of the numerator, he would have a longitudinal mass equal $m$ and a transverse mass equal $m/\gamma$.

Einstein even devoted one of his theoretical works from 1906 to the possibility of experimental measuring the ratio of transverse and longitudinal mass \cite{Einstein 1906}. Unfortunately, Einstein used the ordinary and wrong formula with longitudinal mass (like Bucherer in 1904 \cite{Bucherer 1904}):
\begin{equation}
\label{mistake}
E_{k}=\frac{\mu_{\parallel}\,\mathrm{v}^{2}}{2}\:\:\: \left(\approx \frac{m \mathrm{v}^2}{2}+\frac{3}{4}\frac{m\mathrm{v}^4}{c^2}+...\right) \:\:\: \text{(incorrect)}.
\end{equation}

Despite that, as well as (\ref{Einstein masses}) and (\ref{Lorentz masses}), the presented theory of masses was called by Einstein as Lorentz--Einstein theory \cite{Einstein 1906}, which is somewhat justified \cite{Wroblewski historia}. 
However, writing the formula (\ref{mistake}) is surprising because already in 1905 Einstein gave the correct expression for work $W$ of load acceleration (equals kinetic energy) \cite{Einstein June}:
\begin{equation}
E_k=W=mc^{2}(\gamma-1)\approx \frac{m \mathrm{v}^2}{2}+\frac{3}{8}\frac{m\mathrm{v}^4}{c^2}+... \ .
\end{equation}
Because of the facts presented above Einstein never accepted the idea of mass depending on velocity \cite{Hecht}. Evidence of that is a letter to Barnett from 1948 whose fragment can be found in Okun's work \cite{Okun}, and its corrected translation in discussion \cite{putting rest}. Many critics of relativistic mass refer to this letter. But they forget that from the beginning Einstein was having problems defining mass and he was searching for the right definition. Surprisingly, he did not find it in the mass-energy relation nor in the work from 1905 \cite{Einstein 1905}, nor in the work from 1907 \cite{Einstein 1907}, where he included the following equations (see also \cite{Field}):
 \begin{equation}
\label{E=mc2}
m=\frac{E_0}{c^2} \ \ \ , \ \ \ E=\frac{m c^2}{\sqrt{1-\mathrm{v}^2/c^2}},
\end{equation}
where: $m$ -- rest mass, $E_0$ -- rest energy, $E$ -- relativistic energy (total energy of the body, means the sum of kinetic and resting energy).

Another step of the evolution of the relativistic equation of motion was rewriting Einstein equation (\ref{Einstein dynamics})  by Planck in 1906 \cite{Planck} in the form:
\begin{equation}
\label{Planck simple}
m \gamma \mathbf{a}=q(\mathbf{E}+\mathbf{v}\times\mathbf{B})-q(\mathbf{v}\mathbf{E})\mathbf{v}/c^{2}.
\end{equation}

Because of the last element in this equation, it did not have the simple form, so Planck gave his equation the rate form:
\begin{equation}
\label{Planck rate}
q(\mathbf{E}+\mathbf{v}\times\mathbf{B})=:\mathbf{F}=\frac{d\mathbf{p}}{dt}=\frac{d(m\gamma \mathbf{v})}{dt}.
\end{equation}
This equation could be presented seemingly in the simple form:
\begin{equation}
\label{apparent simple form}
\mathbf{F}=m\frac{d(\gamma \mathbf{v})}{dt}=m\frac{d\vec{\nu}}{dt},
\end{equation}
where $\vec{\nu}=\gamma \mathbf{v}$ is the three-dimensional part of the four-velocity.
But from the perspective of 3D such an equation does not have correctly separated mass and acceleration (in the sense of (\ref{Newton simple}) or (\ref{Lorentz masses})). In the following sections we will also see that this apparently simple form does not translate into an equation with a variable mass. 

\section{V. The mass of Lewis-Tolman 1908--1912 and the Feynman Lectures 1963--2011}
Until that time, mass in the relativistic theory had been mostly electromagnetic mass, and its transformation principles resulted from Maxwell equations and the formula for Lorentz force of motion equations. It was only in 1908 Lewis and 1909 when Lewis and Tolman generalized the Einstein's relation of equivalency of rest mass and energy (\ref{E=mc2}) onto a case of motion \cite{Lewis 1908, Lewis}:
\begin{equation}
\label {M=E/c2}
M=\frac{E}{c^2},
\end{equation}
where: $M$ -- relativistic mass also described by (\ref{relativistic mass}).
Originally, the authors used $m$ instead of $M$ (and rest mass was denoted with $m_0$ instead of $m$). Lewis showed in \cite{Lewis 1908} that the principle of mass and energy equivalence (\ref{M=E/c2}) implies a relativistic mass formula (\ref{relativistic mass}). The same derivation is given
in the Feynman Lectures on Physics \cite{Feynman}(Sec. 15-9), what is copied in work \cite{Field 2018}.

In 1912, Tolman introduced theoretical dependence of mass from velocity based on the relativity principle \cite{Tolman}:
\begin{equation}
\label{relativistic mass}
M=m\gamma=\frac{m}{\sqrt{1-\mathrm{v}^2/c^2}}=\mu_{\perp}.
\end{equation}
Tolman considered the principle of momentum conservation in a perfect inelastic collision. 
 Its derivation which regards mass as a coefficient of velocity in the expression for momentum determined transverse mass in accordance with (\ref{Abraham}). But earlier Lewis and Tolman had stressed a more universal, i.e. (\ref{M=E/c2}), character of mass $M$.  Contrary to the popular belief, it is not Einstein who should be considered the author of the relativistic mass but Lewis and Tolman \cite{Hecht}. Einstein did not consider this concept until 1946, but he did not explicitly accept it \cite{Einstein 1946}.

In 1914, the ideas of Lewis and Tolman were independently confirmed in a more comprehensive and a bit forgotten work by Lorentz \cite{Lorentz 1914}. Lorentz used the same notations $M$ and $m$ as in this work. He also called mass $M$ as transverse mass, but he did not use longitudinal mass in his work.

Derivations such as that by Tolman for relativistic mass are propagated in several excellent textbooks, including the famous Feynman’s lectures \cite{Feynman} (Sec. 16-4). Unlike Tolman, Feynman presented a derivation based on elastic collision. In section 15-1 \cite{Feynman}, Feynman called the relativistic mass the Einstein correction regarding Newtonian mass. It was a logical but not historiographic opinion.

Feynman's lectures are a series of old textbooks from the sixties. However, they are constantly being revised, renewed and authorized. The latest edition has been called the New Millennium Edition (2011) and is authorized by the California Institute of Technology (Caltech). The preface to the previous edition (2005) was written by later Nobel prize winner Kip Thorne. Extensive documentation of corrections to these lectures is available at {\it www.feynmanlectures.caltech.edu/info/}.It is surprising, therefore, that content provided by Feynman and authorized today can be completely independently subjected to such strong criticism (see section VII).

\section{VI. Actual forms of equation of motion}
In 3D formalism, the most popular became the rate form equation by Planck (\ref{Planck rate}), explicitly:
\begin{equation}
\label{Planck formula}
\mathbf{F}=\frac{d}{dt}\left(\frac{m\mathbf{v}}{\sqrt{1-\mathbf{v}^2/c^2}}\right).
\end{equation}
This equation can be presented in the following form:
\begin{equation}
\label{Force}
\mathbf{F}=m\gamma \mathbf{a}+m\gamma^3 (\mathbf{va})\mathbf{v}/c^2=m\gamma \mathbf{a}_{\perp}+m\gamma^3 \mathbf{a}_{\parallel},
\end{equation}
or equivalently using the vector product \cite{Franklin}:
\begin{equation}
\mathbf{F}=m\gamma^3[\mathbf{a}+\mathbf{v}\times(\mathbf{v}\times\mathbf{a})/c^2].
\end{equation}
Somewhat simpler is the counterpart of the equation (\ref{Planck simple}):
\begin{equation}
\label{semi simple}
m\gamma \mathbf{a}=\mathbf{F}-(\mathbf{vF})\mathbf{v}/c^2=\mathbf{F}_{\perp}+\gamma^{-2} \mathbf{F}_{\parallel},
\end{equation}
which gets complicated by using a vector product:
\begin{equation}
m\gamma \mathbf{a}=\gamma^{-2}\mathbf{F}-\mathbf{v}\times(\mathbf{v}\times\mathbf{F})/c^2.
\end{equation}

 In 4D formalism, the movement equation was written in 1908 by Minkowski \cite{Minkowski} basically in one of the following forms:
\begin{equation}
\label{Minkowski}
f^{\mu}=m\frac{d^2 x^{\mu}}{d\tau^2}=m\frac{d \nu^{\mu}}{d\tau}=\frac{d p^{\mu}}{d\tau}=m a^{\mu},
\end{equation}
where: $f^{\mu}$ -- four-vector of force, $\nu^{\mu}$ -- four-vector of velocity, $a^{\mu}$ -- four-vector of acceleration, $\tau$ -- rest time of body. The price for the simplicity of this equation is redefining of the notions of $f^{\mu}, \nu^{\mu}, a^{\mu}, \tau$ in relation to the 3D notions of $\mathbf{F, v, a}$ and time $t$.
For example, the spatial part of the four-acceleration is expressed as follows:
\begin{equation}
\label{fouracceleration}
\vec{a}=\frac{d\vec{\nu}}{d\tau}=\gamma \frac{d(\gamma \mathbf{v})}{dt}=\gamma^2 \mathbf{a}_{\perp}+\gamma^4 \mathbf{a}_{\parallel}
\end{equation}
and not parallel to ordinary acceleration $\mathbf{a}$.
Thus, the last two expressions in (\ref{Minkowski}) are not the rate form, or the simple form, of the equations of dynamics in the sense understood here. This is evident from the rate form (\ref{Planck rate}), which differs from the above mentioned. Similarly, apparent simple form (\ref{apparent simple form}) differ from (\ref{Minkowski}) gamma factor. The proper simple form has not yet been found, its best substitute being (\ref{semi simple}).

The relativistic equation of movement of a body with variable mass was derived in 1946 by Ackeret \cite{Ackeret}:
\begin{equation}
\label{Ackeret}
\mathrm{F}_{th}=m\gamma^3 \mathrm{a}=\gamma\frac{dm}{dt}\mathrm{u} \ominus \mathrm{v},
\end{equation}
where: $\mathrm{u}$ -- velocity of the jet gasses in external frame, $\ominus$ -- relativistic subtraction of velocity in accordance with (\ref{velocity 0}).
Originally, Ackeret did not use here the thrust force $\mathrm{F}_{th}$, so he reduced one factor $\gamma$ in this equation.
However, his equation is not a full equivalent of Meshchersky formula (\ref{rocket simple}) because it does not contain external force and it determines movement only in one spatial dimension. Similar to other researchers, Ackeret concentrated on a relativistic generalization of the Tsiolkovsky rocket equation.
On the left side of the formula (\ref{Ackeret}) there correctly occurs longitudinal mass; the expression $\gamma dm$ on the right side can be interpreted, with precision to a sign, as relativistic mass of jet gasses (relative to the rocket observer).

In this section of the article it is worth emphasizing that SR has been confirmed experimentally not only in electrodynamics or mechanics, but also at the level of Lorentz invariance in particle physics. For example, the result of  weak interaction research published in 2013 \cite{First test}  and neutrino research results from 2018 \cite{Sudbury} did not confirm any violations of Lorentz symmetry. Despite this, the work from 2019 \cite{Spin 2} shows that in the area of SR (in the context of quantum mechanics of massive gravitons), original theoretical results are still possible. In addition, violations of Lorentz symmetry and CPT are constantly monitored \cite{Kostelecky}.

\section{VII. Adler and Okun offensive 1987--1989}

At the beginning of the 1940s, Landau and Lifshitz released a textbook on the field theory \cite{Landau} which included SR and GR which had no reference to relativistic mass $M$. This trend was continued by Taylor and Wheeler \cite{Taylor} and by Ugarov \cite{Ugarov} in the 1960s.

Let us assume the year 1987 of publication of Adler's work titled ``Does Mass Really Depend on Velocity, Dad?" \cite{Adler} as the formal beginning of open criticism of the notion of relativistic mass. Despite the fact that the work mirrors the past and contemporary views of the majority of physicists (see Meissner's lecture \cite{Meissner}), Adler did not escape certain inconsistencies, crucial for this subject matter, in the formulas for gravitational and inertial mass. Because I belong to the supporters of the notion of relativistic mass, who are in the minority, I concentrate on the mistakes made by its critics.  However, further in the article I also described the errors made by supporters of the relativistic mass in equations (\ref{Wolny}, \ref{converted}, \ref{blad masy}, \ref{blad masy 1}).

Gravitational mass according to Adler had the following form:
\begin{equation}
\label{Adler gravity}
m_g=\frac{m}{\sqrt{1+2\Phi/c^2-\mathrm{v}^2/c^2}}\:\:\:\:\:\:\:\: \text{(incorrect)},
\end{equation}
where: $m_{g}$ -- (passive) gravitational mass, $\Phi$ -- potential of gravitation. Despite the fact that this interesting formula refers to the general theory of relativity, it is contradictory to the velocities near the velocity of light ($\Phi<0$). It implies diminishing of the velocity of light in gravitational field, which is contradictory to the theory of relativity (both SR and GR). 

Adler described inertial mass as follows:
\begin{equation}
\label{Adler inercia}
I=\frac{|\mathbf{F}|}{|\mathbf{a}|}=\frac{m\gamma}{1-(\mathrm{v}/c)^2\cos^2\theta}\:\:\:\:\:\:\: \text{(approximate)},
\end{equation}
where: $I$ -- inertial mass depending on the direction, $\theta$ -- the angle between force and velocity. 
This formula has been corrected in the same parametrization (denoted here $\theta$) by Nowik \cite{Nowik}:
\begin{equation}
I=\frac{m\gamma^3}{\sqrt{\cos^2\theta+\gamma^{4}\sin^2\theta}}=m\gamma\sqrt{\sin^2\alpha+\gamma^4\cos^2\alpha},
\end{equation}
where was introduced additional parametrization of the angle $\alpha$ between acceleration and velocity (see {\bf Fig. \ref{diagram 2}}). It is worth noting that the parametrization of this formula with the angle $\beta$ between force and standard acceleration does not exist. We will later see how it is easiest to fix the formula (\ref{Adler inercia}) by modifying its left-hand side rather than the right-hand side as in (\ref{correct Adler}).  

In 1989, Okun became one of serious critics of the notion of relativistic mass \cite{Okun}. Unfortunately, also he failed to avoid certain mistakes which may point to the biased nature of his article. Okun formulated two questions concerning the relation between energy and mass. First, the most correct formula should be indicated, and secondly, the formula derived by Einstein. The reader could chose from four presented below expressions, which have been paired here according to the correctness, in accordance with the notations adopted in this work (which correspond to the notations in Okun's work):
\begin{equation}
a)\:E_0=Mc^2,\:\:\:\:\:\: b)\:E=mc^2,\:\:\:\:\:\:\: \text{(incorrect)}
\end{equation}
\begin{equation}
c)\:E_0=mc^2,\:\:\:\:\:\:d)\:E=Mc^2.
\end{equation}
According to the author (Okun), twice the answer was to the formally incorrect formula $a)$. While the true formula $d)$ was incorrect for Okun because it was not used by Einstein. In reality, Okun wanted to push forward answer $c)$ but he got lost in their notations. This is proved by the next formula in his article explicitly defining relativistic mass $M$ (which he denoted as $m$ and the rest mass by $m_0$ instead of $m$). In this context, on the last page of his article Okun has objections to Hawking that the only formula he had placed in his ``A Brief History of Time" \cite{Hawking 1} book was equivalent to $d)$ and not $c)$ (although it originally looked like $b)$). It is worth noting that even in 1935 Einstein propagated the formula $c)$ (for $c = 1$), while formula $d)$ actually ignored \cite{Einstein 1935, Flores}.
  
Another element of Okun's work was a formula for the force of gravity, presented without any reference, which was equivalent to:
\begin{equation}
\label{Okun gravitation}
\mathbf{F}_g=-\frac{GM_{\odot}E/c^2}{r^3}[\mathbf{r}+\epsilon\mathbf{v}\times(\mathbf{r}\times\mathbf{v})/c^2]\:\:\:\:\:\:\: \text{(incorrect)},
\end{equation}
for $\epsilon=1$, where: $\mathbf{F}_g$ -- force of gravity, $G$ -- gravitational constant, $M_{\odot}$ -- source of gravitational field, $\mathbf{r}$ -- position vector. This formula for $\epsilon=-1/2$ and $E=E_0$ can be obtained from the transformation of Lorentz force with a precision of second order in $\mathbf{v}$ (obviously, it is not the proper methodology to apply here). Formula (\ref{Okun gravitation}) was uncritically repeated by Roche \cite{Roche} for $\epsilon=1$, despite the fact that the author (R.) quoted the formula for $\mathbf{F}$ and for him ``$\mathbf{i}$" in which $\epsilon=-1$, and $E/c^2$ was replaced with $m\gamma^3$. Okun used his formula for photons, so in fact he applied approximation for low velocities to the ultrarelativistic case. Next, he based on this doubtful approximation one of the pillars of his criticism of the relativistic mass.

Okun's work faced some criticism in collected correspondence \cite{putting rest} which also contained his defence. Okun states there that he had found the formula (\ref{Okun gravitation}) only in one textbook by Bowler \cite{Bowler}. Moreover, referring to E.~Schucking, he presents a more general formula \textit{the relativistic apple} for the gravity force corresponding to (\ref{Okun gravitation}) for small $M_{\odot}$.  In 1998, Okun exchanged emails with M.R. Kleemans – a student of Bernard de Wit, about it \cite{Okun 2}. Both men agreed on the issue of Bowler derivative: \textit{If this is the one you mean, I agree, is not a very good derivation} (Kleemans); and they both had doubts concerning the factor in the second order term.

In 1990, Okun was invited by R.H. Romer to publish in AJP in the context of Adler and Sandin's works. This invitation resulted in a publication \cite{Okun 3} only in 2009 which focused on algebraic aspects of the following relation:
\begin{equation}
m^2=(E/c^2)^2-(\mathbf{p}/c)^2 \ \ \text{or} \ \ E^2=(mc^2)^2+(\mathbf{p}c)^2.
\end{equation}
In addition, Okun once again criticized Hawking for the form of quoting the formula of the mass and energy equivalence in his new book \cite{Hawking 2}.

\section{VIII. Sandin, Penrose defense 1991, 2004}

In 1991, Sandin publishes a work titled ``Defence of Relativistic Mass" \cite{Sandin}. It contains the idea of the following Lemma by Sandin:
\begin{equation}
\mathbf{F}=M\mathrm{(v)}\mathbf{a}+\frac{dM}{dt}\mathbf{v}\:\Rightarrow \:\left(\frac{dM}{dt}\neq 0 \:\:\Rightarrow \:\:M \neq \frac{|\mathbf{F}|}{|\mathbf{a}|} \right),
\end{equation}
whose seemingly trivial proof requires the assumption $M\neq mc^2/v^2$. This Lemma shows that the standard Adler's definition of inertial mass (\ref{Adler inercia}) is not proper if the mass changes. In other words, in the light of this methodology, only transverse mass is fully consistent.

A recognised contemporary relativist, R. Penrose, referred in his book ``The Road to Reality" (2004) in a concise way to the concept of mass \cite{Penrose}. He denoted the relativistic mass $M$ as $m$ and named it \textit{total mass}, and invariable mass $m$ he referred to as $\mu$ (like Einstein) and called it \textit{rest mass}. He described the properties of these types of mass which can be formalized in the notations used here for the isolated frame of two not interacting bodies:
\begin{equation}
M\neq \text{inv} \:\: , \:\: M_1\uplus M_2=M_1+M_2\:\: , \:\: M_1+M_2=\text{const};
\end{equation}
\begin{equation}
m= \text{inv} \:\: , \:\: m_1\uplus_0 m_2\not\equiv m_1+m_2\:\: , \:\: m_1+m_2\not\equiv \text{const};
\end{equation}
where: $\text{inv}$ -- invariant symbol, $\uplus$ -- relativistic mass connection rule, $\uplus_0$ -- rest mass connection rule.
As we can see, mass $M$ and mass $m$ have opposing properties regarding invariance, additivity and conservativity in the additive sense. Due to the lack of additivity of rest mass, the later property can be replaced with:
\begin{equation}
m_1\uplus_0 m_2= \text{const}.
\end{equation}
This formula, however, does not refer to strong interaction. The analysis taking into account the contribution of potential and nuclear energy to mass is included in Einstein's short article \cite{Einstein 1946} or in Brillouin's book \cite{Brillouin}. At present, the principle of mass and energy equivalence in the context of nuclear energy and mass deficit is such a well-established experimental fact that it does not even require a reference. 

\section{IX. Three-dimensional definition of relativistic acceleration}

\begin{figure}[h!]
\centering
\includegraphics[width=9cm]{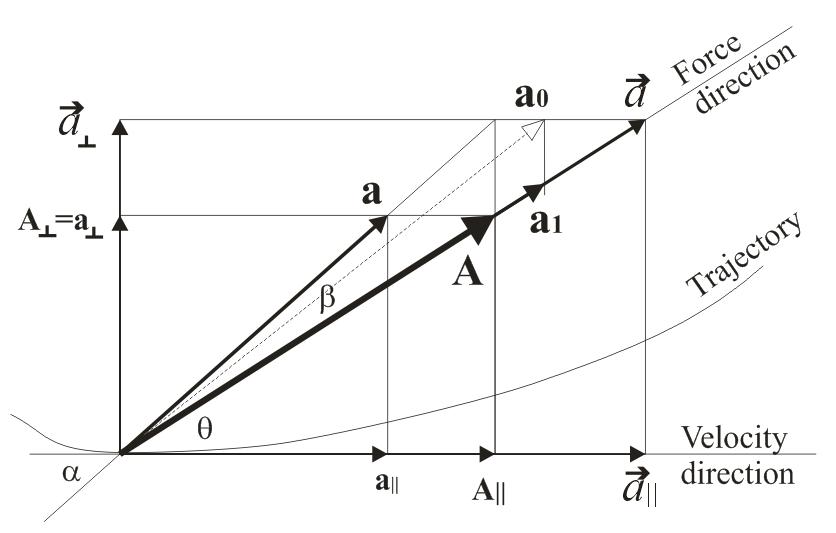}
\caption{Diagram showing the relations of the main acceleration vectors and their components. The new three-dimensional vector of relativistic acceleration $\mathbf{A}$ is a parallel projection of ordinary acceleration $\mathbf{a}$ along the velocity on the direction of force (or spatial part of the four-acceleration $\vec{a}$). Thanks to this, the construction of acceleration $\mathbf{A}$ is simpler than the construction of rest acceleration $\mathbf{a}_0$ and acceleration $\mathbf{a}_1$.}
\label{diagram 2}
\end{figure}

Acceleration is a derivative of velocity with respect to time. In an inertial system, we should use time from this system. And the differential of velocity should be calculated in accordance with the relativistic law of subtracting velocity of type (\ref{velocity 0}), but not exactly (\ref{velocity}). In the plane determined by velocity $\mathbf{v}$ and standard acceleration $\mathbf{a}$ in the natural base of this space, we can define the relativistic acceleration as:

\begin{equation}
\label{A1}
\mathrm{A}_{\parallel}=\frac{(d\mathrm{v}_{\parallel})_{rel}}{dt}=\frac{(\mathrm{v}+d\mathrm{v}_{\parallel})\ominus \mathrm{v}}{dt}=\gamma^2 \mathrm{a}_{\parallel}=\frac{a_{\parallel}}{\gamma^2},
\end{equation}

\begin{equation}
\label{A2}
\mathrm{A}_{\perp}=\frac{(d\mathrm{v}_{\perp})_{rel}}{dt}=\frac{(0+d\mathrm{v}_{\perp})\ominus 0}{dt}=\mathrm{a}_{\perp}=\frac{a_{\perp}}{\gamma^2},
\end{equation}
where: $\mathrm{A}_{\parallel}$, $\mathrm{A}_{\perp}$ -- components of relativistic acceleration, respectively parallel and perpendicular to velocity; $a_{\parallel}$, $a_{\perp}$ -- components of the spatial part of four-acceleration $\vec{a}$. The relativistic differential perpendicular to the velocity does not differ from the ordinary differential, while the parallel differential is calculated as follows:
\begin{equation}
\label{parallel differential}
(\mathrm{v}+d\mathrm{v}_{\parallel})\ominus \mathrm{v}=\frac{\mathrm{v}+d\mathrm{v}_{\parallel} - \mathrm{v}}{1-(\mathrm{v}+d\mathrm{v}_{\parallel})\mathrm{v}/c^2}=\frac{d\mathrm{v}_{\parallel}}{1-\mathrm{v}^2/c^2}+O(d\mathrm{v}^2_{\parallel}).
\end{equation}
By differential definition high order terms are skipped despite using exact equality $=$ instead approximate $\approx$.
The law of velocity subtraction was used in (\ref{A1}, \ref{A2}) in the sense of following the velocity components, and not as a transforming to a rest frame (concerns the second equation). Therefore, strictly speaking, equation (\ref{A2}) does not match to $\ominus_0$ (\ref{velocity}), but is compatible with $\ominus$ (\ref{velocity 0}) in the sense $(\mathbf{u}\ominus_1\mathbf{v})_i := \mathrm{u}_i \ominus \mathrm{v}_i$. The obtained acceleration does not coincide with the rest acceleration (\ref{proper a}) in any component. The above definition implicitly prefers two versors of Frenet natural base: $\hat{\mathbf{t}}$ -- tangential versor, $\hat{\mathbf{n}}$ -- normal versor. Where this base does not exist ($\mathbf{v}=0$ or $\mathbf{a}=0$), any base can be used. In general terms, not all directions produce a correct value of the acceleration component. The third and the most important direction of the conformity of the definition is the direction of the acceleration itself:
 \begin{equation}
\label{A}
\mathrm{A}=\frac{(d\mathrm{v}_{\mathrm{_A}})_{_{rel}}}{dt}=\frac{\mathrm{a}_{\mathrm{_A}}}{1-\mathrm{v}^2_{\mathrm{_A}}/c^2}=\frac{\mathrm{a}\cos \beta}{1-(\mathrm{v}/c)^2\cos^2 \theta},
\end{equation}
where: $\mathrm{v}_{\mathrm{_A}}$ -- velocity component in the direction of $\mathbf{A}$, $\mathrm{a}_{\mathrm{_A}}$ -- standard acceleration component in the direction of $\mathbf{A}$, $\beta$ -- the angle between standard and relativistic acceleration, $\theta$ -- the angle between velocity and relativistic acceleration (also force).
The above relation requires proof based on the components:
\begin{equation}
\frac{\mathbf{a A}/\mathrm{A}}{1-\frac{(\mathbf{vA})^2}{c^2\mathrm{A}^2}}=
\frac{(\gamma^2\mathrm{a}^2_{\parallel}+\mathrm{a}^2_{\perp})\mathrm{A}}{\mathrm{A}^2-\frac{v^2}{c^2}\gamma^4\mathrm{a}^2_{\parallel}}=\sqrt{\gamma^4\mathrm{a}^2_{\parallel}+\mathrm{a}^2_{\perp}}=\mathrm{A}.
\end{equation}
The definition of acceleration $\mathbf{A}$ uses the rule of composition velocity (\ref{velocity 0}) in the direction of the velocity components into vectors $\hat{\mathbf{t}}$, $\hat{\mathbf{n}}$ or $\mathbf{A}$. It turns out that it can be generalised for any direction by introducing a vector rule changing only the parallel component of velocity:
\begin{equation}
\label{velocity parallel}
\mathbf{u}\ominus_{\parallel}\mathbf{v}=\frac{\mathbf{u}-\mathbf{v}+\frac{\mathbf{u}\mathbf{v}}{\mathrm{v}^2}(\mathbf{u}\times\mathbf{v})\times\mathbf{v}/c^{2}}{1-\mathbf{u}\mathbf{v}/c^{2}}=:\mathbf{w}_1.
\end{equation}
Thus, by definition, this operation does not change the component $\mathbf{u}$ perpendicular to $\mathbf{v}$, i.e. $(\mathbf{u}\ominus_{\parallel}\mathbf{v})_{\perp}=\mathbf{u}_{\perp}$, while operation on parallel components is carried out according to (\ref{velocity 0}) and (\ref{velocity}). This subtraction operation introduces the function factor $\varphi$ only for the subtractive vector, as follows $\mathbf{u}\ominus_{\parallel}\mathbf{v}=\mathbf{u}-\varphi\: \mathbf{v}$. The condition for the parallel component is sufficient to calculate the $\varphi$ function. Therefore, the resulting vector $\mathbf{w}_1$ will be called relative axial-type velocity. A similar law of velocity composing was researched by Fern\'andez-Guasti \cite{Mexico} in 2011. Still, his formula ``(1)" is dependent of the coordinate system (like $\ominus_1$), but no specific system was chosen (like Frenet base). But formally the next vector representation ``(6)" is equivalent to (\ref{velocity parallel}). 
Thanks to this law the relativistic differential of velocity vector:
\begin{equation}
\label{dv rel}
(d\mathbf{v})_{rel}=(\mathbf{v}+d\mathbf{v})\ominus_{\parallel}\mathbf{v},
\end{equation}
in the linear part (by differential definition) takes the following form:
\begin{equation}
\label{rozniczka}
(d\mathbf{v})_{rel}=
d\mathbf{v}+\gamma^2\frac{\mathbf{v}(\mathbf{v}d\mathbf{v})}{c^2}=
\gamma^2 d\mathbf{v}_{\parallel}+d\mathbf{v}_{\perp}.
\end{equation}
The operation $\ominus_{\parallel}$ (\ref{velocity parallel}) in velocity differential (\ref{dv rel})  can be equivalently replaced by $\ominus$ (\ref{velocity 0}) or $\ominus_1$ in Frenet base, but not by $\ominus_0$ (\ref{velocity}). A vector definition of relativistic three-acceleration can now be given:
\begin{equation}
\label{correct acceleration}
\mathbf{A}:=\frac{(d\mathbf{v})_{rel}}{dt}:=\lim_{\Delta t\rightarrow 0}\frac{\mathbf{v}(t+\Delta t)\ominus_{\parallel} \mathbf{v}(t)}{\Delta t},
\end{equation}
where $\ominus_{\parallel}$ can be fully equivalently replaced by $\ominus_{\perp}$ (\ref{velocity perp}) or other operations discussed in section XI. In addition, in section XI this definition is presented in the form of a ordinary derivative of relative velocity. The new acceleration corresponds with the four-acceleration and the standard acceleration as follows:
\begin{equation}
\mathbf{A}=\frac{\vec{a}}{\gamma^2}=S_{\mathbf{v}}^{\gamma^2}(\mathbf{a})
=\gamma^2\mathbf{a}_{\parallel}+\mathbf{a}_{\perp},
\end{equation}
where $S_{\mathbf{v}}^{\gamma^2}$ means stretching (directional scaling) in the direction of velocity $\mathbf{v}$ and scale $\gamma^2$. This transformation can be demonstratively interpreted on {\bf Fig. \ref{diagram 2}} as double inversion of Lorentz contraction resulting from the second order differential of position $d^2\mathbf{r}$ which occurs in acceleration. This interpretation would be more accurate for $d\mathbf{r}^2$ instead of $d^2\mathbf{r}$.

For comparison purposes, we will also use the vector law of velocity composing (\ref{velocity}) into velocity differentiation:
\begin{equation}
\label{dv Einstein}
(d\mathbf{v})_0=(\mathbf{v}+d\mathbf{v})\ominus_0 \mathbf{v}=
\gamma d\mathbf{v}+\frac{\gamma^3}{\gamma+1}\frac{\mathbf{v}(\mathbf{v}d\mathbf{v})}{c^2},
\end{equation}
which can be simplified to the following form:
\begin{equation}
\label{dv Einstein 2}
(d\mathbf{v})_0=
\gamma^2 d\mathbf{v}_{\parallel}+\gamma d\mathbf{v}_{\perp},
\end{equation}
This differential may be used to determining the proper (rest) acceleration:
\begin{equation}
\label{proper a}
\mathbf{a}_0=\lim_{\Delta \tau\rightarrow 0}\frac{\mathbf{v}(\tau+\Delta \tau)\ominus_0 \mathbf{v}(\tau)}{\Delta \tau}=\gamma^3\mathbf{a}_{\parallel}+\gamma^2\mathbf{a}_{\perp},
\end{equation}
where the derivative is calculated with respect to the self time $\Delta \tau=\Delta t/\gamma$. The product of resting mass and resting acceleration gives resting force, which gives agreement (\ref{proper a}) with Einstein's masses definition (\ref{Einstein masses}).

It is also worth considering acceleration occurring in (\ref{apparent simple form}):
\begin{equation}
\label{a1}
\mathbf{a}_1:=\frac{d\vec{\nu}}{dt}:=\frac{d(\gamma\mathbf{v})}{dt}:=\lim_{\Delta \tau\rightarrow 0}\frac{\mathbf{v}(\tau+\Delta \tau)\ominus_{\parallel} \mathbf{v}(\tau)}{\Delta \tau}.
\end{equation}
Despite the simplicity of the first two records, this definition has a hybrid nature, because the numerator and denominator refers to different formalisms: four-dimensional or three-dimensional. 
This also applies to the last record, in which the denominator refer to the rest system, and the numerator is based on the new rule of subtracting the components of the velocity, which does not consist in a full transformation to the rest system.
The non-hybrid version of this record is the correct relativistic acceleration (\ref{correct acceleration}) or proper (rest) acceleration (\ref{proper a}). In contrast, the hybrid acceleration itself is related to others as follows:
 \begin{equation}
\mathbf{a}_1=\gamma \mathbf{A}=\vec{a}/\gamma=\gamma^3\mathbf{a}_{\parallel}+\gamma\mathbf{a}_{\perp}.
\end{equation}
The relations between all considered acceleration vectors are shown in the {\bf Fig. \ref{diagram 2}}.

Probably the relativistic velocity differential, for one dimension acceleration, was first calculated by Barrett in 2002 \cite{Barrett}. 
Barrett's differential is only compatible with equation (\ref{parallel differential}) and does not resolve the key difference of vector expressions (\ref{rozniczka}) and (\ref{dv Einstein}). The vector relativistic velocity differential for acceleration was analysed by R\c{e}bilas \cite{Rebilas} in 2008. However, he used the common approach to the algebra of velocity, which led to the confusion which Einstein obtained in (\ref{Einstein masses}). Moreover, R\c{e}bilas applied relativistic addition, not subtraction, of velocity, which forced a rearranged order of operations and an equation inverse of (\ref{dv Einstein}), but equivalent. This differential $(d\mathbf{v})_0$ (\ref{dv Einstein}, \ref{dv Einstein 2}), calculated according to the usual algebra of speed vectors, can also be found in Dragan's studies \cite{Dragan, Dragan AJP}. The square of such rest differential velocity can be seen in the formula of the Lobachevsky-Einstein geometry metric described by Fock in the late fifties, to which Cannoni refers (see Appendix). However, the differential $(d\mathbf{v})_{rel}$ (\ref{dv rel}, \ref{rozniczka}) does not appear in the works of Foc, Barrett, R\c{e}bilas, Dragan, Cannoni or even in the work of Fern\'andez-Guasti containing operation (\ref{velocity parallel}). Ungar \cite{Ungar 1997, Ungar 2007} and Oziewicz \cite{Oziewicz, groupoid, without Lorentz} did not present the differential approach either.

It turns out that Lorentz group (of composition velocity and rotations) has many spectacular properties. The best example is the Thomas precession described in 1926 \cite{Thomas} in the context of centripetal acceleration in quantum mechanics. Thomas precession surprised even Einstein himself \cite{Dragan} and, according to R\c{e}bilas, he surprises physicists to this day \cite{Rebilas 2}. Indeed, the physical phenomenon of Thomas precession in the face of the mathematical effect of Thomas--Wigner rotation gives grounds for an alternative approach to subtraction of velocity vectors \cite{Ungar 1997, Dragan}. Ungar described this as a loop structure (quasi group), which he called the gyrogroup \cite{Ungar 1997, Ungar 2007}. Oziewicz, on the other hand, describes it in terms of relativistic groupoid \cite{groupoid}. 

\section{X. Force and simple second law F=\textit{M}A} 

Planck determined force $\mathbf{F}$ by formula (\ref{Planck rate}) in analogy to Newton formula (\ref{Newton rate}). In Minkowski spacetime, however, we have four-force (\ref{Minkowski}), which is expressed as follows:
\begin{equation}
\label{fourforce}
(f^{\mu})=(f^0, \vec{f} )=(\gamma \mathbf{Fv}/c, \gamma \mathbf{F}).
\end{equation}
Thus, we should know and be able to justify which of the vectors $\mathbf{F}$ or $\vec{f}$ is the vector of force which occurs in the definition of work. The choice of work in the form of $\delta W=\mathbf{F}\delta \mathbf{r}$ is the standard SR, and choosing $\delta w=\vec{f} \delta \mathbf{r}$ leads to alternative energy \cite{Osiak}. It is also worth knowing that “four-dimensional work” nullifies itself $f_{\mu}\delta x^{\mu}=0$. For this and other reasons (see (\ref{work-impulse}) for the first component), alternative energy is incorrect and inconsistent with experiment. The norm of four-force in the signature $(+,-,-,-)$ is:
\begin{equation}
\label{norm}
f_{\mu}f^{\mu}=-\mathbf{F}_{\parallel}^2-\gamma^2 \mathbf{F}_{\perp}^2.
\end{equation}
Since the norm of four-force is equal with accuracy to the sign to the norm of three-force parallel to velocity, this can suggest that vector $\mathbf{F}$ is more important than $\vec{f}$. But let us consider the following four-covector of work-impulse:
\begin{equation}
\label{work-impulse}
(\delta W_{\mu})=(f_{\mu})\delta \tau=(\mathbf{F}\delta \mathbf{r}/c, -\mathbf{F}\delta t),
\end{equation}
where $\delta W_0=\delta W/c$, but $\delta W_t=\delta W$. Exchange $\mathbf{F}$ to $\vec{f}$ would destroy transformational laws for (\ref{work-impulse}). In the work \cite{Franklin}, the vector $\mathbf{F}$ is called the Lorentz force, and the vector $\vec{f}$ or the four-vector $f^{\mu}$ is the Minkowski force, while the product of $m\mathbf{a}$ is called Newtonian force. One of the objectives of present  work is to unify Lorentz force with properly understood Newtonian force.

The derivation of the second law of dynamics in a simple form will begin 
with the correction Adler's formula (\ref{Adler inercia}). It turns out that this formula should be presented in the sense of component of the force direction:
\begin{equation}
\label{correct Adler}
\frac{|\mathbf{F}|}{\mathrm{a}_{\mathrm{_F}}}=\frac{\mathrm{F}}{\mathrm{a} \cos \beta}=\frac{m\gamma}{1-(\mathrm{v}/c)^2\cos^2\theta}.
\end{equation}
Which can be transformed to an equation in the direction of force operation:
\begin{equation}
\label{in force direction}
\mathrm{F}=m \gamma \frac{\mathrm{a}_{\mathrm{_F}}}{1-\mathrm{v}_{\mathrm{_F}}^2/c^2},
\end{equation}
which can be proved with the help of (\ref{A}) using the parallelism of $\mathbf{F}$ and $\mathbf{A}$ from (\ref{Force}) and (\ref{A1}, \ref{A2}). Using (\ref{A}), we can write it in the following form:
\begin{equation}
\mathrm{F}=m \gamma \frac{(d\mathrm{v}_{\mathrm{_F}})_{_{rel}}}{dt}=m \gamma \frac{(\mathrm{v_{_F}}+d\mathrm{v_{_F}})\ominus \mathrm{v_{_F}}}{dt}.
\end{equation}
This equation is true for any force direction, regardless of the direction of velocity. Also true are the equations in the direction parallel and perpendicular to velocity:
\begin{equation}
\mathrm{F}_{\parallel}=m \gamma \frac{(d\mathrm{v}_{\parallel})_{rel}}{dt} \:\:\:\: , \:\:\:\: 
\mathrm{F}_{\perp}=m \gamma \frac{(d\mathrm{v}_{\perp})_{rel}}{dt}.
\end{equation}
It turns out that the correction of Adler formula (\ref{in force direction}) will not be true for every direction. For example, in the direction $\mathbf{a}$ a slightly different relation occurs:
\begin{equation}
\label{in a direction}
\mathrm{a}=\frac{\mathrm{F}_{\mathrm{a}}}{m \gamma(1+\gamma^2\mathrm{v}_{\mathrm{a}}^2/c^2)}.
\end{equation}
However, thanks to (\ref{dv rel}) in every direction and in every orthonormal Cartesian coordinate system the motion equation is true in the following vector form:
\begin{equation}
\label{F=MA}
\mathbf{F}=m \gamma \frac{(d\mathbf{v})_{rel}}{dt}=M \mathbf{A}.
\end{equation} 
This is the title relativistic motion equation in the simple form. It is worth stressing that for the basic directions, i.e. parallel and perpendicular to velocity and parallel to force (or relativistic acceleration), the equation does not go beyond the principle (\ref{velocity 0}). Only other arbitrary directions require the use of the equation (\ref{velocity parallel}) or (\ref{velocity perp}, \ref{antisymmetric}, \ref{fourspeed sub}, \ref{Ungar sub}) (see below).

At the end of this section, consider the form of a motion equation that will appeal to the supporters of the rest mass as an inertial mass:
\begin{equation}
\mathbf{F}=m \frac{d(\gamma\mathbf{v})}{dt}=m \frac{(\mathbf{v}+d\mathbf{v})\ominus_{\parallel} \mathbf{v}}{d\tau}=m \mathbf{a}_1.
\end{equation}
However, acceleration $\mathbf{a}_1$ has an artificial character, which was already justified and will still be in section XIII.

\section{XI. Variable mass motion equation
and velocities subtraction methods}

Now a relativistic generalization of the Meshchersky equation (\ref{rocket simple}), more general than the Ackeret equation (\ref{Ackeret}), will be found. In order to do that, we will use four-vector rate form of the motion equation for a system of body and additional mass:
\begin{equation}
f_{ext}^{\mu}=\frac{d\left[ m(\tau)\dot{x}^{\mu}+\delta m(\tau) \upsilon^{\mu}\right]}{d\tau},
\end{equation}
where: $f_{ext}^{\mu}$ -- external four-fource, $m(\tau)$ -- body rest mass depends on self time, $\dot{x}^{\mu}$ -- body four-velocity as a derivative of the position with respect to self time, $\upsilon^{\mu}$ -- four-velocity of little additional mass $\delta m$.
Calculation of the derivative leads to the equation: 
\begin{equation}
\label{rocket 2}
f_{ext}^{\mu}=m\ddot{x}^{\mu}+\dot{m}\dot{x}^{\mu}+\delta \dot{m} \upsilon^{\mu}.
\end{equation}
The condition of orthogonality $f_{ext}^{\nu}\dot{x}_{\nu}=0$ allows to calculate:
\begin{equation}
\delta \dot{m}=-\dot{m} c^2/(\upsilon^{\nu}\dot{x}_{\nu}).
\end{equation}
Application of this relation in (\ref{rocket 2}) will lead to a ready equation:
\begin{equation}
\label{covariant}
f_{ext}^{\mu}=m\ddot{x}^{\mu}+\dot{m}[\dot{x}^{\mu}-\upsilon^{\mu}c^2/(\upsilon^{\nu}\dot{x}_{\nu})].
\end{equation}
Note that the expression in square brackets is orthogonal to the four-velocity $\dot{x}^{\mu}=\nu^{\mu}$. It turns out that this expression with the opposite sign equals Oziewicz's binary relative velocity \cite{binary} -- here, for example,  the velocity of jet gases relative to the rocket:
\begin{equation}
\label{Oziewicz}
\omega^{\mu}(\nu, \upsilon)=\frac{c^2}{\upsilon\cdot\nu} \ \upsilon^{\mu}-\nu^{\mu} \ \ (\text{Oziewicz's}) .
\end{equation}
Relative velocity is a space-like four-vector, which is normalized like three-dimensional velocity in rocket frame (with accuracy to the sign) $\omega_{\mu}\omega^{\mu}=-(\mathbf{u}\ominus_0\mathbf{v})^2=-\mathbf{u}_0^2$, not like four-velocities $\upsilon_{\mu}\upsilon^{\mu}=\nu_{\mu}\nu^{\mu}=c^2$. If there is a relative velocity in the equation, it should also be visible in a three-dimensional version analogous to the Meshchersky or Ackeret equation:
\begin{equation}
\label{general}
\mathbf{F}_{ext}=m\frac{d(\gamma\mathbf{v})}{dt}-\gamma\frac{dm}{dt}\mathbf{u}\ominus_{\perp}\mathbf{v}.
\end{equation}
The operation $\ominus_{\perp}$ is a new, but the simplest so far, way to subtract velocity vectors:
 \begin{equation}
\label{velocity perp}
\mathbf{u}\ominus_{\perp}\mathbf{v}=\frac{\mathbf{u}-\mathbf{v}+(\mathbf{u}\times\mathbf{v})\times\mathbf{v}/c^{2}}{1-\mathbf{u}\mathbf{v}/c^{2}}=:\mathbf{w}.
\end{equation}
This subtraction determines the relative velocity of the jet gases, therefore $\mathbf{w}$ it will be called the jet-type relative velocity. The operation $\ominus_{\perp}$ can be interpreted on the basis of a simplified method of its derivation, that is, by projecting the quasi-four-vector $(c,\mathbf{v})$ onto a hyperplane perpendicular to it in the direction of the quasi-four-vector $(c,\mathbf{u})$. The projection result contains the analyzed formula $(...,-\mathbf{u}\ominus_{\perp}\mathbf{v})$. However, in three-dimensional terms, it is enough to know that $\mathbf{u}\ominus_{\perp}\mathbf{v}=\psi \: \mathbf{u}-\mathbf{v}$, and the compatibility of parallel components (up to $\mathbf{v}$) with (15) or (16) enables the calculation of the function $\psi$.
That is why the rule $\ominus_{\perp}$ (\ref{velocity perp}) of velocity used for parallel velocities, similar to $\ominus_{\parallel}$ (\ref{velocity parallel}), coincides with the original rule $\ominus_0$ (\ref{velocity}). All three rules work differently for perpendicular velocities, $\ominus_{\parallel}$ does not change the perpendicular component, and $\ominus_{\perp}$ changes it the most. In spite of this difference, the modified versions of the velocity substraction formula are differentially equivalent:
\begin{equation}
\label{div}
(\mathbf{v}+d\mathbf{v})\ominus_{\perp}\mathbf{v}=(\mathbf{v}+d\mathbf{v})\ominus_{\parallel}\mathbf{v}=(d\mathbf{v})_{rel},
\end{equation}
where the details are presented in (\ref{rozniczka}), and differ from (\ref{dv Einstein}). Operation $\ominus_{\perp}$ results from the analysis of the variable mass equation, but it can also be derived from Oziewicz's binary relative velocity. It is enough to show that this spatial-like four-velocity can be expressed using three-velocity $\mathbf{w}$ in a way analogous to four-force (\ref{fourforce}). Then, such velocity $\mathbf{w}$ determines the subtraction of the velocities $\mathbf{u}$ and $\mathbf{v}$ in another equivalent way:
\begin{equation}
\label{Oziewicz 2}
(\omega^{\mu})=(\omega^0, \vec{\omega} )=(\gamma \mathbf{wv}/c, \gamma \mathbf{w})\ \ \rightarrow \ \ \mathbf{w}=\mathbf{u} \ominus_{\perp}\mathbf{v},
\end{equation}
where $\gamma$ depends on $\mathbf{v}$. The vector and value of three-dimensional relative jet-type velocity $\mathbf{w}$ are not equal to the vector and value $\mathbf{u}_0$ of velocity $\mathbf{u}$ in a system moving at velocity $\mathbf{v}$, but it is a form of its transformation. After all, it is obvious that for $\mathbf{v} \neq 0$ the input reference system is not resting for $\mathbf{v}$. In other words, we are talking here about the additional relativity of relative velocity associated with the third reference system (our own). Contrary to intuition, this double kind of relativity also occurs in Galileo spacetime. The idea of such relative relativity is qualitatively similar to the idea of Oziewicz ternary relative velocity \cite{Oziewicz}. By referring to such a ``binary-ternary" velocity in a given reference system to the velocity of the body relative to the uniform co-moving system $\mathbf{w}(t_0, t)=\mathbf{v}(t)\ominus_{\perp}\mathbf{v}(t_0)$, one can simply express relativistic acceleration:
\begin{equation}
\label{simple derivativ}
\mathbf{A}(t_0)=\frac{d\mathbf{w}}{dt}(t_0, t)|_{t=t_0}.
\end{equation}
Instead of the jet-type velocity $\mathbf{w}$, the axial-type velocity $\mathbf{w}_1$ (\ref{velocity parallel}) or $\mathbf{W}$ (\ref{antisymmetric}) can be used here, equivalently.
An analogous definition for relative four-velocity leads to the ordinary four-acceleration:
\begin{equation}
a^{\mu}(\tau_0)=\frac{d\omega^{\mu}}{d\tau}(\tau_0, \tau)|_{\tau=\tau_0},
\end{equation}
where self time was used.

An alternative approach to the algebra of velocity quasigroup (or groupoid) is not something completely new and appears in the literature on the subject \cite{Oziewicz, Mexico, Dragan, Ungar 2007}. 
Particularly interesting is the antisymmetric operation from Dragan's monograph containing lectures from SR \cite{Dragan}:
\begin{equation}
\label{antisymmetric}
\mathbf{u}\ominus_{\wedge}\mathbf{v}=\frac{(\gamma^{-1}_{\mathrm{u}}+\gamma^{-1}_{\mathrm{v}})(\gamma_{\mathrm{u}}\mathbf{u}-\gamma_{\mathrm{v}}\mathbf{v})}
{1-\frac{\mathbf{u}\mathbf{v}}{c^2}+\gamma_{\mathrm{u}}\gamma_{\mathrm{v}}(1-\frac{\mathrm{u}^2 \mathrm{v}^2}{c^4})}=:\mathbf{W} \ \ (\text{Dragan's}).
\end{equation}
This complicated formula is the solution to the physically simple problem of finding the velocity $\mathbf{W}$, which converts the velocity $\mathbf{u}$ to the velocity $\mathbf{v}$, namely $\mathbf{u}\ominus_0 \mathbf{W} = \mathbf{v}$. In other words, the relative velocity  $\mathbf{W}$ of systems in which the velocities of the same body are $\mathbf{u}$ and $\mathbf{v}$, respectively, is sought. So the solution to this reverse velocity issue is the new subtraction formula $\mathbf{W}=\mathbf{u}\ominus_{\wedge} \mathbf{v}$. The asymmetry of this operation ($\mathbf{u}\ominus_{\wedge} \mathbf{v}=-\mathbf{v}\ominus_{\wedge} \mathbf{u}$) results from the properties of Lorentz transformation. Due to this asymmetry and interpretation of relative velocity, it cannot be ruled out that operation $\ominus_{\wedge}$ (\ref{antisymmetric}) is a more reasonable subtraction of velocities than operation $\ominus_0$ (\ref{velocity}).
It turns out that Dragan's velocity is a special case of ternary relative velocity introduced by Oziewicz in 2004 (see (\ref{binary subtraction})), which can be expressed covariantly by three four-velocities:
\begin{equation}
\label{Oziewicz ternary}
\xi^{\mu}=\frac{\sigma\cdot(\upsilon+ \nu)(c^2\delta^{\mu}_{\nu}-\sigma^{\mu}\sigma_{\nu})(\upsilon^{\nu}-\nu^{\nu})}{(\sigma\cdot\upsilon)^2+(\sigma\cdot\nu)^2+c^2\upsilon\cdot\nu -c^4},
\end{equation}
where $\sigma^{\mu}$ is the four-velocity reference system in which the velocity $\upsilon^{\mu}$ relative to $\nu^{\mu}$ is determined. Ternary velocity is an antisymmetric space-like four-vector, whose special case is binary velocity:
\begin{equation}
\label{ternary binary}
\xi^{\mu}(\sigma, \nu, \upsilon)=-\xi^{\mu}(\sigma, \upsilon, \nu)\ \ , \ \ \xi^{\mu}(\nu, \nu, \upsilon)=\omega^{\mu}(\nu, \upsilon).
\end{equation}
Ternary velocity is based on the Lorentz transformation in the same way as Dragan's velocity, to which it is reduced in the reference system $\sigma^{\mu}=(c, 0, 0, 0)$:
\begin{equation}
\label{Oziewicz vs Dragan}
(\xi^{\mu})_{\{\sigma\}}=(\xi^0, \vec{\xi})_{\{\sigma\}}=(0, \mathbf{W})_{\{\sigma\}} \ \ \rightarrow \ \ \vec{\xi}_{\{\sigma\}}=\mathbf{W}.
\end{equation}
The similarity of 4D and 3D formulas is clearer if the ternary velocity is expressed by subtraction of binary velocities -- what was done by Oziewicz \cite{binary, Oziewicz}:
\begin{equation}
\label{binary subtraction}
    \xi^{\mu}=\frac{(\gamma_1+\gamma_2)(\gamma_2\omega_2^{\mu}-\gamma_1\omega_1^{\mu})}{\gamma_1^2+\gamma_2^2+\gamma_1\gamma_2(1+\omega_1\cdot\omega_2/c^2)-1} \ \ (\text{Oziewicz's}),
\end{equation}
where: $\omega^{\mu}_1=\omega^{\mu}(\sigma,\nu)$, $\omega^{\mu}_2=\omega^{\mu}(\sigma,\upsilon)$, $\gamma_1=(1+\omega_1^2/c^2)^{-1/2}$. Contrary to appearances, gamma factors do not differ by the plus sign from Lorentz factors ($\omega^2<0$), but they are formally invariants here.

Despite the complexity and differences, it can be proved that operation $\ominus_{\wedge}$ (\ref{antisymmetric}) is differentially equivalent to operations $\ominus_{\parallel}$ (\ref{velocity parallel}), $\ominus_{\perp}$ (\ref{velocity perp}):
\begin{equation}
\label{div Dragan}
(\mathbf{v}+d\mathbf{v})\ominus_{\wedge}\mathbf{v}=(\mathbf{v}+d\mathbf{v})\ominus_{\parallel}\mathbf{v}=(d\mathbf{v})_{rel}.
\end{equation}
Therefore, under the interpretation of operation $\ominus_{\wedge}$ the velocity differential  is explicitly equal to the Lorentz transformation velocity $\mathbf{W}=(d\mathbf{v})_{rel}$. This velocity tends to zero when calculating the derivative defining acceleration $\mathbf{A}$. Accordingly, acceleration $\mathbf{A}$ is defined in a system that asymptotically tends to the main system, and not to any other, in particular (not) to the resting system of the described body (for $\mathrm{v} \neq 0$). Even the four-acceleration, which uses the body's resting time, does not meet this criterion.

Based on the proove of (\ref{div Dragan}), many simpler antisymmetric and differential-equivalent operations can be constructed. The simplest of them is operation that looks almost like a four-velocity substraction:
\begin{equation}
\label{fourspeed sub}
\mathbf{u}\ominus_{\vee}\mathbf{v}=\gamma^{-1}_{\mathbf{u}\mathbf{v}}(\gamma_{\mathrm{u}}\mathbf{u}-\gamma_{\mathrm{v}}\mathbf{v}), \ \ \ 
\gamma^{-1}_{\mathbf{u}\mathbf{v}}=\sqrt{1-\mathbf{u}\mathbf{v}/c^2}.
\end{equation}
In his work, Ungar called a similar operation Einstein's cooperation \cite{Ungar 2007}:
\begin{equation}
\label{Ungar sub}
\mathbf{u}\boxminus \mathbf{v}=2 \cdot \frac{\gamma_{\mathrm{u}}\mathbf{u}-\gamma_{\mathrm{v}}\mathbf{v}}{\gamma_{\mathrm{u}}+\gamma_{\mathrm{v}}},  \ \ \ (\text{Ungar's})
\end{equation}
whereby the specific original multiplication $\otimes$ has been replaced here by a simple multiplication ``$\cdot$" by 2. Both operations $\ominus_{\vee}$ (\ref{fourspeed sub}) and $\boxminus$ (\ref{Ungar sub}) can be interpreted as antisymmetric subtraction (as simple as possible) of spatial parts of four-velocities. However, in (\ref{fourspeed sub}) the gamma factor of the result contains $\mathbf{uv}$ instead of the square of velocity, while in (\ref{Ungar sub}) it is the arithmetic mean of the gamma factors for $\mathrm{u}$ and $\mathrm{v}$. The differential still unifies the new and the previous operations (except for operation $\ominus_0$):
\begin{equation}
\label{div Ungar}
(\mathbf{v}+d\mathbf{v})\boxminus\mathbf{v}=(\mathbf{v}+d\mathbf{v})\ominus_{\vee}\mathbf{v}=(\mathbf{v}+d\mathbf{v})\ominus_{\parallel}\mathbf{v}=(d\mathbf{v})_{rel}.
\end{equation}
In view of these next equivalences, the interpretation of examples (\ref{fourspeed sub}) and (\ref{Ungar sub}) is no longer so important. Strong interpretation of operation $\ominus_{\wedge}$ (\ref{antisymmetric}) is sufficient, additionally supported by interpretations of operations $\ominus_{\parallel}$ (\ref{velocity parallel}) and $\ominus_{\perp}$ (\ref{velocity perp}). All the operations listed above ($\ominus_{\wedge}, \ominus_{\parallel}, \ominus_{\perp},\ominus_{\vee},  \boxminus$) are differentially equivalent. This means that they are an important class of local operations on the four-velocity hyperboloid $\nu^{\mu}\nu_{\mu}=c^2$, to which vector operation $\ominus_0$ (\ref{velocity}) does not belong. The four-velocity hyperboloid is a curved surface, so the important role of differentials in the context of such differential geometry should not come as a surprise (see Appendix).

The most general variable mass motion equation (\ref{general}) can be written in a simple form as follows:
\begin{equation}
\label{full general}
m \gamma \mathbf{A}=m\frac{d(\gamma \mathbf{v})}{dt}=\mathbf{F}_{ext}+\mathbf{F}_{th}=\mathbf{F}_{ext}+\frac{\gamma dm}{dt}\mathbf{u}\ominus_{\perp}\mathbf{v},
\end{equation}
or in a less expanded form:
\begin{equation}
\label{full general 1}
M \mathbf{A}=\mathbf{F}_{ext}+\frac{\partial M}{\partial t}\mathbf{u}\ominus_{\perp}\mathbf{v}.
\end{equation}
Operation $\ominus_{\perp}$ here refers to finite velocities, so it cannot be replaced by another differential equivalent operation. In this way, this simplest operation stands out from the others. Despite being based on Oziewicz binary velocity, this simple operation has not yet appeared in the literature.

Equation (\ref{full general 1}), like operation $\ominus_{\perp}$, does not appear in the literature, either. For example, consider the equation given in 2019 by Wolny and Strza\l{}ka \cite{Wolny}:
\begin{equation}
\label{Wolny}
\frac{d\mathbf{p}}{dt}=\mathbf{F}_{ext}+\frac{dM }{dt} \mathbf{u}=\mathbf{F}_{ext}+\gamma \frac{dm }{dt} \mathbf{u}+m \frac{d\gamma}{dt} \mathbf{u} \ \ \ \text{(incorrect).}
\end{equation}
For $dm/dt=0$, this equation is contrary to the ordinary motion equation (\ref{Planck rate}) due to the last non-disappearing term.
The authors tried to generalize the equation (\ref{semi simple}) for the variable rest mass. Work \cite{Wolny} does not contain an exact derivation of the relativistic issue of the variable rest mass,
but loose generalizations of cases $dm/dt=0$ and $\mathbf{u}=0$ modeled on a non-relativistic (or relativistic) version. In one spatial dimension, equation (\ref{Wolny}) can be converted equivalently to the following form:
\begin{equation}
\label{converted}
m\gamma^3\frac{d\mathrm{v}}{dt}=\frac{\mathrm{F}_{ext}}{1-\mathrm{uv/c^2}}+\gamma \frac{dm }{dt} \mathrm{u}\ominus \mathrm{v} \ \ \ \text{(incorrect).}
\end{equation}
Nevertherest, the authors gave a different equation \cite{Wolny}:
\begin{equation}
\label{Wolny 2}
m\gamma^3\frac{d\mathrm{v}}{dt}=\mathrm{F}_{ext}+\gamma \frac{dm }{dt} \mathrm{u}\ominus \mathrm{v}.
\end{equation}
This spatially one-dimensional equation is correct and consistent with equation (\ref{full general 1}). Omitting the contradiction of derivation with (\ref{Wolny}) and (\ref{converted}), the above equation (\ref{Wolny 2}) could be considered as a partial generalization of the Ackeret equation (\ref{Ackeret}) for external force. However, deriving the full equation (\ref{full general 1}) for all cases and spatial dimensions was much more difficult.

\section{XII. Inertial mass definitions}

After World War II (and after the atomic bombings of Hiroshima and Nagasaki), Einstein wrote a short and rarely cited popular science article about the analysis of the generalization of the principle of mass and energy equivalence in the case of potential gravity energy and nuclear energy \cite{Einstein 1946}. It was basically Einstein's only article in which he considered mass in the general form $E/c^2$, and not only in resting form $E_0/c^2$ (\cite{Einstein 1946} vs \cite{Hecht}). Unfortunately, Einstein tried to reject this concept on the basis of a negligible mass of heat energy. However, in the case of nuclear energy, its contribution to mass is already measurable. Einstein described this fact, but without explicit reference to the expression $E/c^2$ (or $E_0/c^2$).

As mentioned earlier, Einstein in his letter to Barnett from 1948 postulated the need to present a good definition of the mass dependent on the velocity. Below are given 10 equivalent, but not identical, definitions of inertial mass (dependent on velocity), of which the first 3 are sufficiently general. 
\begin{enumerate}
\item Force and relativistic acceleration ratio (for $\mathrm{A}\neq 0$):
\begin{equation}
\label{1}
M:=\frac{\mathrm{F}}{\mathrm{A}}=\frac{\mathrm{F}}{\frac{(d\mathrm{v_{_F}})_{_{rel}}}{dt}}.
\end{equation}
The direction of the force is absolutely free, and the velocity subtraction rule for differential of velocity does not go beyond (\ref{velocity 0}). 
\item Time component of mass-momentum four-vector:
\begin{equation}
\label{pt}
M:=p^t\:\:\:,\:\:\:\hat{p}=m \gamma \frac{\partial}{\partial t}+m\gamma \mathbf{v} \frac{\partial}{\partial \mathbf{r}},
\end{equation}
where $\hat{p}$ is expressed here in the language of modern differential geometry (classical, not quantum).
This definition of mass does not refer directly to the velocity of light $c$ and it is correct also in the Galilean spacetime.
\item Mass energy equivalent in general form:
\begin{equation}
M:=\frac{E}{c^2}=\frac{p_t}{c^2}\:\:\:,\:\:\:\check{p}=m \gamma c^2 dt+m\gamma \mathbf{v} d\mathbf{r}.
\end{equation}
This definition is not the same as (\ref{pt}) because energy-momentum four-covector $\check{p}$ does not exist in Galilean spacetime, which makes the correspondence with a non-relativistic theory difficult. Einstein used the formula $E_0=mc^2$ or $E=m\gamma c^2$, but he eventually did not decide to adopt the formula $E=Mc^2$ \cite{Hecht, Einstein 1907, Einstein 1935, Flores}. While there is no rational reason to consider this formula improper \cite{Brillouin, Einstein 1946, Criado, Rindler 2}. Most probably, Lewis was the first ones to propose it \cite{Lewis 1908}. Feynman presented a similar approach \cite{Feynman} (Sec. 15-9). The general version of the mass-energy equivalence principle has such great heuristic power that, despite some subtleties, it can be studied theoretically \cite{Gravitational mass}, as well as planned experimental studies \cite{Savrov} of the phenomenon of gravitational mass deficit.
\item Momentum and velocity ratio (for $\mathrm{v}\neq 0$):
\begin{equation}
M:=\frac{\mathrm{p}}{\mathrm{v}}.
\end{equation}
This is the simplest definition of the notion of mass, but it may not be convincing in the light of the redefinition of the formula for momentum with the factor $\gamma$. This definition was used, among others, by Abraham, Lewis and Tolman and Feynman. 
\item Generalization of Abraham's first formula:
\begin{equation}
\label{5}
M:=\frac{d\mathrm{p}}{(d\mathrm{v})_{rel}}.
\end{equation}
The original formula (\ref{Abraham}) without the relativistic differential of velocity determined longitudinal mass. 
\item Generalization of the Kaufmann  formula:
\begin{equation}
\label{6}
M:=\frac{1}{\mathrm{v}}\frac{d E}{(d\mathrm{v})_{rel}}.
\end{equation}
This definition used without the velocity substration formula (\ref{Kaufmann}) determined the longitudinal mass.
\item Generalization of Abraham's second formula:
\begin{equation}
\label{7}
M:=\frac{d^2 E}{(d\mathrm{v})^2_{rel}}.
\end{equation}
In its original version (\ref{Abraham two}) for Lagrangian function and without $rel$, this formula defines longitudinal mass. 
\item Thrust mass implies with (\ref{Ackeret}) or (\ref{full general}):
\begin{equation}
M_{\delta m}:=\gamma |\delta m|.
\end{equation}
It can be noticed that a change in the rest mass leads to relativistic mass regardless of the direction and value of jet velocity with the Lorentz factor for body velocity.
\item A simple definition consistent with the Sandin lemma:
\begin{equation}
M(\mathrm{v})=\text{const} \:\: \wedge \:\: \mathrm{a} \neq 0  \:\:\Rightarrow \:\:M:= \frac{\mathrm{F}}{\mathrm{a}}=\frac{\mathrm{F}_{\perp}}{\mathrm{a}_{\perp}}.
\end{equation}
Inertial mass favours transverse mass as it does not change during measuring (defining).
\item Average longitudinal mass:
\begin{equation}\label{10.0}
M:=\frac{1}{\mathrm{v}}\int_0^{\mathrm{v}}\mu_{\parallel}(\mathrm{v}) d\mathrm{v}.
\end{equation}
This definition allows us to easily calculate acceleration time with a constant force:
\begin{equation}
F\int_0^t dt=\int_0^{\mathrm{v}}\mu_{\parallel}(\mathrm{v}) d\mathrm{v}\ \ \Rightarrow  \ \ t=\frac{M\mathrm{v}}{\mathrm{F}},
\end{equation}
which approaches $\infty$ for $\mathrm{v}\rightarrow c$. And if we make the mistake of averaging the relativistic mass (instead of the longitudinal mass):
\begin{equation}
\label{blad masy}
F\int_0^T dt=\int_0^{\mathrm{c}}M(\mathrm{v}) d\mathrm{v}\ \ \ (\text{incorrect}),
\end{equation}
this time will turn out to be finite \cite{Gluza}:
\begin{equation}
\label{blad masy 1}
T=\frac{\pi}{2}\frac{mc}{\mathrm{F}} \ \ \ (\text{incorrect}).
\end{equation}
The formula (\ref{blad masy}) can be easily corrected by changing $d\mathrm{v}$ to $(d\mathrm{v})_{rel}$. This observation allows to rewrite (\ref{10.0}) as a self-consistent formula for averaging one type of mass:
\begin{equation}
\label{10.1}
M(\mathrm{v})\equiv \frac{1}{\mathrm{v}}\int_0^{\mathrm{v}}M(\mathrm{u}) (d\mathrm{u})_{rel}.
\end{equation}
\end{enumerate}
All the definitions lead to the same formula $M=m\gamma$. In some cases it is obvious and in others it requires elementary differentiation or integration.

The time has come to present the final interpretation of mass $M$ depending on velocity. First and foremost, it is not self or rest mass but the mass expressing the dynamic relation of motion of a body in relation to the spacetime. As postulated by Mach, physical mass is not only an immanent property of a body but it also determines its relations with the environment. In this case, it is the relation connected with motion, which is relative. Mach criticized the substantial definition of mass by Newton and in 1883 he wrote: \textit{The true definition of mass can be deduced only from the dynamical relations of bodies} \cite{Mach}. An increase of mass along with velocity is similar to the contraction of length, but it operates in the opposite direction. In a rest frame, a body has a particular rest length and rest mass. In a frame moving fast relative to a body, the instantaneous length of the body is smaller and the inertial mass is higher. Contraction of Lorentz length is considered to be an experimental and theoretical fact, but the increase of mass has many opponents among physicists (e.g. \cite{Oas, Meissner, Okun, Adler, Nowik}). Supporters of the relativistic mass are rather an older generation of relativists (e.g. \cite{Sandin, Lorentz 1914, Brillouin, Feynman, Penrose, Rindler 1}).
Mass growth can also be compared to time dilation. It is often acknowledged that the relativistic mass results from time dilation and is denied its own sense. The matter, however, is not obvious, because the acceleration determining the inertial mass is the second (and not the first) derivative of displacement over time. Time is subject to dilation, and displacement is subject to a form of contraction, so with the second derivative it is difficult to say that only the first effect decides. 
These controversies arose from the lack of a solid definition of inertial mass based on the relativistic law of velocity subtraction, ergo on the lack of a correct manner of determining acceleration. The lack of these definitions entailed numerous, and not always correct, speculations. The dependence of mass on velocity has been demonstrated experimentally by Kaufmann four years before Einstein announced his theory of relativity. The fact of this dependence was undeniable, and only its detailed formula was debatable (Lorentz's formula versus Abraham's formula). In later years the analogous experiments \cite{Bucherer} were repeated but the results of the measuring points were presented in the form of rest mass $m=m_0(\mathrm{v})=const$. In the author's opinion, such an approach is inconsistent with the historic Kaufmann and Newton’s point of view (in the sense of its simple form), although it was convenient in the verification of the theory. The point is that the calculation of the rest mass $m$ of a moving body requires the application and selection of a specific theory (e.g. Lorentz or Abraham) and knowledge of the value of $c$. In contrast, the value of $M$ in the Kaufmann experiment results more directly from the measurement under the foundations of Newtonian dynamics, which are not disputed by the theory of relativity.

This work concerns only Special Relativity (SR), so it is not attempting to prove that $M$ is a passive or active gravitational mass, because it would require the use of General Relativity (GR). This does not mean, however, that this is not true, which is partly suggested by the work from 2019 \cite{Gravitational mass}, contrary to some previously cited works. Similarly, the relativistic mass of photons has not been discussed here, because it would require the use of quantum mechanics and the formulae of wave-particle duality $E=\mathrm{hf}$ and $\mathrm{p}=\mathrm{h}/\lambda$. The use of these formulae for photons can be found in another author's work \cite{Koczan}. The particle mass with classic spin \cite{Kosyakov} was also not discussed here. It turns out that for such particles there is some variation already at the level of rest mass definition $m=\sqrt{p_{\mu}p^{\mu}}/c$ or $\tilde{m}=p_{\mu}\nu^{\mu}/c^2$. The most precision determination of the rest mass of an electron based on the measurement and calculations of quantum electrodynamics is in the works \cite{Sturm, Zatorski}.

\section{XIII. Spacetime-space corespondence}

Let's consider in Minkowski space the physical quantity: four-vector $(q^{\mu})=(q^0,\vec{q})$ or scalar $\mathrm{q}$. This quantity corresponds to the three dimensional vector $\mathbf{Q}$ or a semiscalar $Q$. If the following relation occurs:
\begin{equation}
\label{rank}
\vec{q}=\gamma^n \mathbf{Q} \ \ \ \text{or} \ \ \ \mathrm{q}=\gamma^n Q 
\end{equation}
we will say that the given quantity has the rank $n$. The {\bf Tab. \ref{quantities}} presents seven most important mechanical quantities (four kinetic and three dynamic), which meet the aforementioned condition.

\begin{table}[h!]
\centering
\caption{Corespondence of 4D and 3D quantities. The observable physical world is three-dimensional, so rules are needed to define three-dimensional observables based on spacetime quantities. Theory of relativity does not explain why time is not observed as a geometrical dimension. Perhaps this problem will be explained by quantum gravity \cite{Rovelli, ilusion}.}
\begin{ruledtabular}
\begin{tabular}{l l l l}
Quantity & Spacetime (4D) & Rank  & Space (3D)\\
\hline	
Position & $(x^{\mu})=(ct, \mathbf{r})$ & \ \ 0 & $\mathbf{r}=\vec{x}$\\
Velocity & $(\nu^{\mu})=(\gamma c, \gamma \mathbf{v})$ & \ \ 1 & $\mathbf{v}=\vec{\nu}/\gamma$\\
Rel. velocity & $(\omega^{\mu})=(\gamma \mathbf{wv}/c, \gamma \mathbf{w})$ & \ \ 1 &$\mathbf{w}=\vec{\omega}/\gamma$\\
Acceleration & $(a^{\mu})=(\gamma^4\mathbf{av}/c, \vec{a})$ & \ \ 2 & $\mathbf{A}=\vec{a}/\gamma^2$\\
            & $a^{\mu}=\frac{d\nu^{\mu}}{d\tau}=\frac{d\omega^{\mu}}{d\tau}$ &  & $\mathbf{A}=\frac{d\mathbf{w}}{dt}$ \\
            & $\vec{a}=\gamma^2\mathbf{a}+\gamma^4(\mathbf{av})\mathbf{v}/c^2$ &  &  \\ 
Momentum & $(p^{\mu})=(m\gamma c, m\gamma \mathbf{v})$ & \ \ 0 & $\mathbf{p}=\vec{p}$\\
Force & $(f^{\mu})=(\gamma\mathbf{Fv} /c, \gamma \mathbf{F})$ & \ \ 1 & $\mathbf{F}=\vec{f}/\gamma$\\
Mass & $m=\sqrt{p_{\mu}p^{\mu}}/c$ & \  --1 & $M=m \gamma$\\
& $M=p^t=p^0/c=m\gamma$ &  & $M=p^t$\\
\end{tabular}
\end{ruledtabular}
\label{quantities}
\end{table}

We can see that without introducing a new definition of relativistic acceleration $\mathbf{A}$ and relativistic mass $M$, the {\bf Tab. \ref{quantities}} would not be so clear. Velocity is the first derivative with respect to time, so it has the rank 1 (both ordinary and relative). By analogy, acceleration has the rank 2. Force is a derivative of momentum (rank 0) with respect to time, so it has rank 1. The consequence of the relation of momentum and velocity or force and acceleration is mass rank of $-1$. The second rank of acceleration $\mathbf{A}$ is a counter-argument against the acceleration $\mathbf{a}_1$ of the first rank as a seemingly elementary relativistic acceleration. Therefore, the four acceleration can be expressed as follows:
\begin{equation}
(a^{\mu})=(a^0,\ \vec{a})=(\gamma^2\mathbf{Av}/c,\ \gamma^2\mathbf{A}).
\end{equation}

We can see that there are two natural conventions of mass. The first one is based on the scalar $m$ in 4D and semiscalar $M$ in 3D. Definition (\ref{rank}) refers to this approach. The second convention regards the time component of four-momentum (\ref{pt}), which in both spaces gives mass $M$. This approach corresponds very well with mass in Galilean spacetime.  

In the context of this section, it is worth knowing that there are three-vector invariants of Lorentz transformation described by R\c{e}bilas \cite{Rebilas 3}.

\section{XIV. Conclusion}

The title purpose of this work to write a relativistic equation of motion in simple form $\mathbf{F}=M \mathbf{A}$ has been fully realised by the equation (\ref{F=MA}). This was possible thanks to deriving original definitions (\ref{correct acceleration}) and (\ref{simple derivativ}) of relativistic acceleration $\mathbf{A}$ based on the relativistic differential of velocity $(d\mathbf{v})_{rel}$ (\ref{dv rel}, \ref{rozniczka}, \ref{div}, \ref{div Dragan}) or axial-type, jet-type and binary and ternary relative velocities (\ref{velocity parallel}, \ref{velocity perp}, \ref{Oziewicz 2}, \ref{antisymmetric}, \ref{Oziewicz vs Dragan}):
\begin{equation}
\label{conclusion}
\mathbf{A}:=\frac{(d\mathbf{v})_{rel}}{dt}:=\frac{d\mathbf{w}_1}{dt}:=\frac{d\mathbf{w}}{dt}:=\frac{d\mathbf{W}}{dt}\equiv\frac{d(\gamma \mathbf{v})}{\gamma dt}.
\end{equation}
 The relativistic differential definition of the main directions $\parallel$ (\ref{A1}), $\perp$ (\ref{A2}), $\mathbf{F}$ or $\mathbf{A}$ (\ref{A}) does not exceed beyond unidimensional velocity composition formula $\ominus$ (\ref{velocity 0}) in the sense of following to velocity components and not as a transforming to a rest frame (vector operation $\ominus_1$). Most importantly, the acceleration $\mathbf{A}$ is parallel to the force $\mathbf{F}$ and has the rank of 2.

But for any give direction, the velocity vector operation $\ominus_0$ (\ref{velocity}) and even components operation $\ominus$ (\ref{velocity 0}) (vector operation $\ominus_1$) must be replaced with  operation $\ominus_{\parallel}$ (\ref{velocity parallel}) or operation $\ominus_{\perp}$ (\ref{velocity perp}) or operation $\ominus_{\wedge}$ (\ref{antisymmetric}), equivalently. The penultimate operation $\ominus_{\perp}$ was derived with a completely independent method, which increases the importance of those differential equivalent operations (\ref{div}, \ref{div Dragan}). This independent method is in fact the derivation of a general relativistic equation (in SR) of body motion with a variable rest mass in the covariant form (\ref{covariant}), three dimensional form  (\ref{general}) and even the simple form (\ref{full general}).
Operations $\ominus_{\parallel}$ and $\ominus_{\perp}$ turned out to be differentially-equivalent to the antisymmetric operation $\ominus_{\wedge}$ found in the literature. Operation $\ominus_{\wedge}$ allows the velocity differential to be interpreted asymptotically within a local observer, without any reference to the instantaneous resting system of the body. In addition, the result of $\ominus_{\wedge}$ turned out to be a three-dimensional version of the covariant 4D Oziewicz ternary relative velocity. A simplified differential-equivalent version $\ominus_{\vee}$ of the found operation was also given. Ultimately, it turned out that all considered alternative operations $\ominus_{\parallel}$, $\ominus_{\perp}$, $\ominus_{\wedge}$, $\ominus_{\vee}$ are differentially-equivalent to operation $\boxminus$ similar to Einstein's cooperation in the theory of gyrogroup.

The consequence of the correct definition of relativistic three dimensional acceleration, motion equation in the simple form and of using the relativistic velocity substruction formula is clarification, generalisation and unification of numerous formulae for inertial mass (relativistic mass). Five of the ten definition mass formulae (\ref{1}, \ref{5}, \ref{6}, \ref{7}, \ref{10.1}) given in section XII are based on the notion of relativistic differential velocity. In these definitions, it is not necessary to replace the standard subtraction velocity $\ominus$ (\ref{velocity 0}) with other alternative operations considered (and it would be equivalent anyway).
 In other words, in the direction parallel to the velocity, all the laws of velocity subtraction considered in the work are differential-equivalent, without exceptions. Similarly, the defined inertial mass meets the general principle of mass and energy equivalence $E=Mc^2$, without exception for kinetic energy. In addition, it has been shown at the level of dimensional analysis and transformation laws that the time component of the four-momentum is strictly speaking the mass $p^t=M$, and not energy or its other conversion equivalent (e.g. $E/c$). This fact also applies to Galileo's spacetime, where $p^t=M=m$. The energy in Minkowski spacetime is strictly the time component of the four-momentum covector $p_t=E$, which has no equivalent in Galileo's spacetime. The analysis of the equation describing relativistic rockets showed that jet gases have a relativistic mass with a Lorentz factor for the rocket. 
And the relative velocity of jet gases itself is determined by the operation $\ominus_{\perp}$, by subtraction rocket velocity.
The 4D covariant form of jet gases velocity relative to the rocket is exactly Oziewicz binary relative velocity.

The final confirmation of the correctness of the adopted definitions is the simplification of the correspondence between the 4D spacetime and 3D space quantities described in the previous section. The introduction of acceleration $\mathbf{A}$ significantly simplified and unified this correspondence. An important element of the justification of this correspondence was previously introduced work-impulse four-covector $\delta W_{\mu}$ (\ref{work-impulse}). This object determined correspondence for force regardless of Planck's formula (\ref{Planck formula}). 

\section{Appendix: Metrics in Velocity Space}


Only two non-trivial velocity differentials are considered in this work. The first of these is the rest differential $(d\mathbf{v})_0$ (\ref{dv Einstein}, \ref{dv Einstein 2}) relates to the system, which instantaneous co-moving with the body. The second is a relativistic differential $(d\mathbf{v})_{rel}$ (\ref{dv rel}, \ref{rozniczka}). The relativistic differential defines the relativistic acceleration and the spatial part of the four-acceleration. However, the rest differential can only be used to determine the rest acceleration.


The square of velocity differential creates a kind of non-Euclidean metric in space of velocity. Fock called this geometry (for rest differential) ``The Lobachevsky--Einstein Velocity Space" \cite{Fock}. Often this space is called hyperbolic geometry
for Special Relativity \cite{Barrett, Cannoni, Ungar 1997, Ungar 2007}. It has a metric that is equal to the square of the rest differential \cite{Cannoni, Fock}:
\begin{equation}
\label{metric 0}
(d\mathbf{v})^2_0=\frac{(d\mathbf{v})^2-(\mathbf{v}\times d\mathbf{v})^2/c^2}{(1-\mathrm{v}^2/c^2)^2}=
\gamma^4 d\mathbf{v}^2_{\parallel}+\gamma^2 d\mathbf{v}^2_{\perp}.
\end{equation}
It is not difficult to notice that the metric is overstated at least $\gamma^2$ times. Indeed, this is not a three-velocity $\mathbf{v}$ metric, but a four-velocity $\nu^{\mu}=(\gamma c, \gamma \mathbf{v})$ metric on hyperboloid $\nu^2=c^2$:
\begin{equation}
\label{metric 1}
(d\mathbf{v})^2_0=-d\nu_{\mu}d\nu^{\mu}=-d\omega_{\mu}d\omega^{\mu}.
\end{equation}
At the same time, as you can see, it is also the metric for binary velocity differential. This result should not be surprising, 
because the binary velocity $d\omega^{\mu}$ of four-velocity $\nu^{\mu}+d\nu^{\mu}$ relative $\nu^{\mu}$ refers to the temporarily resting system (for $\nu^{\mu}$).

In view of the above, the 3D velocity metric should refer only to the three-dimensional part of the four-velocity metric, i.e. $(d\vec{\nu})^2=d(\gamma \mathbf{v})^2$, and should also be scaled by factor $\gamma^2$:
\begin{equation}
\label{metric rel}
(d\mathbf{v})^2_{rel}=\frac{(d\vec{\nu})^2}{\gamma^2}=\left(\frac{d(\gamma \mathbf{v})}{\gamma}\right)^2.
\end{equation}
Differentiation is not commutative with multiplication and division by the factor $\gamma$, so we get a result consistent with (\ref{rozniczka}):
\begin{equation}
\label{metric rel 1}
(d\mathbf{v})^2_{rel}=
\gamma^4 d\mathbf{v}^2_{\parallel}+d\mathbf{v}^2_{\perp}.
\end{equation}
However, one may ask why at the beginning we did not take the usual $d\mathbf{v}^2$ metric for three-dimensional velocity. This did not happen because {\it a priori} we do not know the metric for $d\mathbf{v}$, and we know the metric for both $d\nu^{\mu}$ and $d\vec{\nu}$. So we had to use a known metric, but we had to relate it to 3D velocity, not a four-velocity that was generally $\gamma$ times greater.

If we determine the differential ternary velocity in the selected reference system with the four-velocity $\sigma^{\mu}(\vec{0})=(c,\vec{0})$:
\begin{equation}
\label{dxi}
d\xi^{\mu}=\xi^{\mu}\left(\sigma(\vec{0}),\nu,\nu+d\nu\right),
\end{equation}
it will be equal to the ternary velocity differential understood strictly as:
\begin{equation}
\label{dxi 1}
d\xi^{\mu}=d\tau_0\frac{d\xi^{\mu}}{d\tau}\left(\sigma(\vec{0}),\nu(\tau_0),\nu(\tau)\right)\Big|_{\tau=\tau_0}.
\end{equation}
Now the relativistic velocity metric can be written by the formula:
\begin{equation}
(d\mathbf{v})^2_{rel}=-d\xi_{\mu}d\xi^{\mu}.
\end{equation}
In a sense, this reflects the covariant nature of the relativistic differential of velocity and its metric, while at the same time showing its ternary character and the need to refer to a given reference system. In Special Relativity, both 3D velocity and its differential depend on the reference system.

\section{Acknowledgments}

Many thanks to my friend engineer Marcin Nolbrzak, without whose this article would not have came into being. Our discussions have deepened the analysis of the Kaufmann experiment measuring the relativistic mass.  

Also big thanks go to Jacek Zatorski in his technical assistance in the process of typesetting and language corrections.

I also would like to thank to my supervisor, Pawe\l{} Kozakiewicz thanks to whom I began to publish science articles. Similarly, I thank Jerzy Kami\'nski, who encouraged me to publish in the best journals. In turn, I thank Tomasz Borowski for motivational support in the difficult publication process.

I thank my student Juliusz Ziomek for his questions concerning the variable mass motion equations.

Thanks very much to Ryszard Kostecki for the many constructive comments.

Finally, thank you to Zbigniew Oziewicz from Mexico for consultations on binary and ternary relative velocity.


\section{Conflict of Interest}
The author declares no conflict of interest.

This work is in line with Einstein's Special Relativity and the Lorentz group. However, it contains important and novelty issues commonly ignored in the context of velocity, acceleration, mass, and equations of motion.

\section{List of Symbols}
\begin{longtable}{ll}
  $\alpha$ &  the value of the angle between ordinary  \\ & acceleration and velocity \\
  $\mathrm{a}$ &  the value of ordinary  acceleration\\
  $\mathrm{a}_{\parallel}$ & component (coordinate) of ordinary \\ & acceleration parallel to velocity\\
  $\mathrm{a}_{\perp}$ & component (coordinate) of ordinary \\ & acceleration perpendicular to velocity\\
   $\mathrm{a}_{\mathrm{A}}$ & component of ordinary acceleration\\ & in the direction of relativistic acceleration\\
   $\mathrm{a}_{\mathrm{F}}$ & component of ordinary acceleration\\ & in the direction of force (equal to $\mathrm{a}_{\mathrm{A}}$)\\
  $\mathbf{a}$ &  the 3D vector of ordinary  acceleration\\
  $\mathbf{a}_{\parallel}$ & vector component of ordinary \\ & acceleration parallel to velocity\\
  $\mathbf{a}_{\perp}$ & vector component of ordinary \\ & acceleration perpendicular to velocity\\
  $|\mathbf{a}|$ &  value of ordinary  acceleration (equal to $\mathrm{a}$)\\
  $\mathbf{a}_0$ &  the rest 3D vector of acceleration\\
  $\mathbf{a}_1$ &  ratio of the force vector and rest mass\\
  $a^{\mu}$ &  four-acceleration 4D vector\\
  $a^0$ &  c-time ($ct$) component of four-acceleration\\
  $\vec{a}$ &  the spatial 3D part of four-acceleration\\
  $\vec{a}_{\parallel}$ & vector component of $\vec{a}$ parallel to velocity\\
  $\vec{a}_{\perp}$ & vector component of $\vec{a}$ perpendicular to velocity\\
  $a_{\parallel}$ & component (coordinate) of $\vec{a}$ parallel to velocity\\
  $a_{\perp}$ & component (coordinate) of $\vec{a}$ perpendicular to \\ & velocity\\
   $\mathrm{A}$ &  the value of relativistic  acceleration\\
  $\mathrm{A}_{\parallel}$ & component (coordinate) of relativistic \\ & acceleration parallel to velocity\\
  $\mathrm{A}_{\perp}$ & component (coordinate) of relativistic \\ & acceleration perpendicular to velocity\\
  $\mathbf{A}$ &  the 3D vector of relativistic  acceleration\\
 $\mathbf{A}_{\parallel}$ & vector component of relativistic \\ & acceleration parallel to velocity\\
 $\mathbf{A}_{\perp}$ & vector component of relativistic \\ & acceleration perpendicular to velocity\\
  $\beta$ & the value of the angle between ordinary  \\
  & acceleration and force\\
   $\mathrm{B}$ & value of the magnetic field induction vector \\
   $\mathbf{B}$ & magnetic field induction vector\\
   $c$ & speed of light in a vacuum\\
   $\delta$ & capacitor plates distance\\
$\delta^{\mu}_{\nu}$ & Kronecker delta (equal to 1 or 0)\\
$\delta m$ & small attaching (or detaching) rest mass\\
$\delta w$ & abstract work of four-force\\
$\delta W$ & real work (portion)\\
$\delta W_{\mu}$ & work-force impuls 4D four-covector\\
$\delta W_0$ & c-time ($ct$) component of $\delta W_{\mu}$\\
$\delta W_t$ & time component of $\delta W_{\mu}$ (equal to work)\\
$\Delta m$ & rest mass increment (implicitly negative)\\
$\frac{\partial}{\partial \mathbf{r}}$ & three base vectors (versors) for spatial $x, y, z$\\ & coordinates in the derivative representation \\
$\frac{\partial}{\partial t}$ & base vector (versor) for time in the derivative \\ & operator representation \\ 
$\frac{\partial M}{\partial t}$ & partial derivative of relativistic mass with \\ & respect to time regarding  changing rest mass\\
$\frac{dM}{dt}$ & total derivative of relativistic mass with \\ & respect to time\\

$d\nu^{\mu}$ & differential of four-velocity\\
$d\vec{\nu}$ & differential of 3D part of four-velocity\\
$(d\vec{\nu})^2$ & Euclidean square of 3D differential\\
$d\mathrm{v}$ &  ordinary differential of the value of velocity \\
& (identity equal to $d\mathrm{v}_{\parallel}$)\\
$d\mathrm{v}_{\parallel}$ & component (coordinate) of ordinary  differential\\ & of velocity parallel to velocity\\
$d\mathrm{v}_{\perp}$ & component (coordinate) of ordinary differential\\ & of velocity perpendicular to velocity\\
$d\mathrm{v}_{\mathrm{F}}$ & component (coordinate) of ordinary differential\\ & of velocity in force direction\\
$(d\mathrm{v})_{rel}$ & relativistic differential of value of velocity\\
& (identity equal to $(d\mathrm{v}_{\parallel})_{rel}$)\\
$(d\mathrm{v}_{\parallel})_{rel}$ & component (coordinate) of relativistic velocity \\ & differential parallel to velocity\\
$(d\mathrm{v}_{\perp})_{rel}$ & component (coordinate) of relativistic velocity\\ & differential perpendicular to velocity\\
$(d\mathrm{v}_{\mathrm{A}})_{rel}$ & component (coordinate) of relativistic velocity\\ & differential in direction of $\mathbf{A}$\\
$(d\mathrm{v}_{\mathrm{F}})_{rel}$ & component (coordinate) of relativistic velocity\\ & differential in direction of $\mathbf{F}$ (same for $\mathbf{A}$)\\
$d\mathbf{v}$ &  ordinary differential of velocity vector \\
$d\mathbf{v}_{\parallel}$ & vector component of ordinary  differential\\ & of velocity parallel to velocity\\
$d\mathbf{v}_{\perp}$ & vector component of ordinary differential\\ & of velocity perpendicular to velocity\\
$(d\mathbf{v})_0$ & rest differential of velocity vector in instantaneous\\ & co-moving system\\
$(d\mathbf{v})_{rel}$ & relativistic differential of velocity vector\\
$(d\mathbf{v})^2_0$ & invariant rest metric in velocity space\\ 
$(d\mathbf{v})^2_{rel}$ & relativistic metric in velocity space\\
$d\omega^{\mu}$ & binary velocity of $\nu^{\mu}+d\nu^{\mu}$ relative to $\nu^{\mu}$\\
& (equal differential of $\omega^{\mu}$)\\
$\frac{d\mathbf{w}}{dt}$ & derivative of jet-type velocity $\mathbf{w}(t_0, t)$ \\ & with respect to $t$ (default for $t=t_0$)\\
$\frac{d\mathbf{w}_1}{dt}$ & derivative of axial-type velocity of the body \\ & relative to  uniform co-moving system respect \\ & to time (default at the point of overlapping)\\
$\frac{d\mathbf{W}}{dt}$ & derivative of ternary-type velocity of the body \\ & relative to  uniform co-moving system respect \\ & to time (default at the point of overlapping)\\
$d\xi^{\mu}$ & ternary velocity of $\nu^{\mu}+d\nu^{\mu}$ relative to $\nu^{\mu}$\\
& in reference to stationary system $\sigma^{\mu}(\vec{0})$\\
& (equal differential of $\xi^{\mu}$ in stationary system)\\

$\epsilon$ & factor equal to $\pm1$ or $-1/2$ \\
$e$ & absolute value of the electron charge\\
$E$ & total relativistic energy\\
$E_0$ & rest energy\\
$E_k$ & kinetic energy\\
$\mathbf{E}$ & electric field intensity vector\\
$\mathbf{E}_{\parallel}$ & electric field parallel to velocity \\
$\mathbf{E}_{\perp}$ & electric field perpendicular to velocity\\
$\varphi$ & factor-function for subtrahend velocity in $\ominus_{\parallel}$\\ 
$\Phi$ & potential of gravitation\\
$\mathrm{f}$ & photon frequency\\
$f^{\mu}$ & four-force 4D vector\\
$f^0$ & c-time ($ct$) component of four-force\\
$\vec{f}$ & spatial 3D part of four-force\\
$\mathrm{F}$ & value of force vector\\
$\mathrm{F}_{\parallel}$ & force component (coordinate) \\ & parallel to velocity\\
$\mathrm{F}_{\perp}$ & force component (coordinate) \\ & perpendicular to velocity\\
$\mathrm{F}_{\mathrm{a}}$ & force component (coordinate) \\ & in direction of $\mathbf{a}$\\
$\mathrm{F'}_{\parallel}$ & rest value of force component\\ & parallel to velocity\\
$\mathrm{F'}_{\perp}$ & rest value of force component\\ & perpendicular to velocity\\
$\mathbf{F}$ & force 3D vector\\
$\mathbf{F}_{\parallel}$ & force parallel to velocity\\
$\mathbf{F}_{\perp}$ & force perpendicular to velocity\\
$|\mathbf{F}|$ & values of force (equal to $\mathrm{F}$)\\
$\mathbf{F}_{\delta m}$ & force acting on jet gases\\
$\mathbf{F}_{ext}$ & external force\\
$\mathbf{F}_{g}$ & gravitational force\\
$\mathbf{F}_{th}$ & thrust force\\
$\mathbf{F}_{tot}$ & total force\\
$\gamma$ & Lorentz factor\\
$\gamma_{\mathrm{u}}$ & Lorentz factor for speed $\mathrm{u}$\\
$\gamma_{\mathrm{v}}$ & Lorentz factor for speed $\mathrm{v}$\\
$\gamma_{\mathbf{uv}}$ & Lorentz type factor with the square of velocity \\ & converted to $\mathbf{uv}$\\
$\gamma_1$ & invariant Lorentz type factor for binary relative \\ & velocity  between $\nu^{\mu}$ and  $\sigma^{\mu}$ (Oziewicz gamma factor)\\
$\gamma_2$ & invariant Lorentz type factor for binary relative \\ & velocity between $\upsilon^{\mu}$ and  $\sigma^{\mu}$ (Oziewicz gamma factor)\\
$G$ & gravitational constant\\
$\mathrm{h}$ & Planck constant\\
$h$ & capacitor height\\
$h'$ & height (length) of the electric field in the capacitor\\
$_i$ & spatial index for 3D velocity (without metric minus)\\
$\mathbf{i}$ & Roche inertial mass reference\\
$I$ & Adler inertial mass\\
$\lambda$ & photon wavelength\\
$L$ & Lagrangian\\
$\mu$ & mass in Kaufmann's theoretical formula (longitudinal,\\ & although he measured transverse)\\
$\mu_{\parallel}$ & longitudinal mass\\
$\mu_{\perp}$ & transverse mass\\
$\mu'_{\perp}$ & transverse mass according to Einstein\\
$^\mu$ & upper contravariant index of four-vector\\
$_\mu$ & lower covariant index of four-covector\\
$m$ & rest invariant mass\\
$\tilde{m}$ & Kosyakov rest mass\\
$m_0$ & rest mass symbol in references\\
$m_g$ & Adler gravitational mass\\
$\dot{m}$ & derivative of rest mass with respect to self time\\
$m_1$ & rest mass of the first body of binary system\\
$m_2$ & rest mass of the second body of binary system\\
$M$ & relativistic inertial mass (total mass)\\
$M_{\delta m}$ & relativistic mass of jet gases\\
$M_{\odot}$ & gravitational field source mass (e.g. Sun's)\\
$M_1$ & relativistic mass of the first body of binary system\\
$M_2$ & relativistic mass of the second body of binary system\\
$(M/e)'$ & recompiled relativistic mass and charge ratio\\

$^\nu$ & upper contravariant index of four-vector\\
$_\nu$ & lower covariant index of four-covector\\
$\hat{\mathbf{n}}$ & normal versor of Frenet base\\ 
$\vec{0}$ & null 3D velocity vector\\ 
$\ominus$ & one-dimensional operation of relativistic \\ & velocities subtraction\\
$\ominus_0$ & Einstein's vector operation of relativistic \\ & velocities subtraction\\
$\ominus_1$ & vector operation of relativistic velocities subtraction,\\
& based on $\ominus$ in the distinguished Cartesian system\\
& (equal to $\ominus_{\parallel}$ in the Frenet base)\\
$\ominus_{\parallel}$ & vector operation of relativistic velocities subtraction,\\
& (which only changes direction of subtracted velocity)\\
$\ominus_{\perp}$ & vector operation of relativistic velocities subtraction,\\
& based on the binary velocity of rocket jet gases\\
$\ominus_{\wedge}$ & Dragan antisymmetric vector operation of relativistic \\ & velocities subtraction, leading to ternary velocity\\
$\ominus_{\vee}$ & the simplest differential equivalent of antisymmetric \\ & vector operation of relativistic velocities subtraction\\
$\boxminus$ & Ungar's operation called Einstein's cooperation\\
& for antisymmetric subtraction of velocity vectors\\ 
$\oplus$ & relativistic Einstein rule of velocity vectors addition\\
& (asymmetric, nonassociative, arbitrary order)\\
$\uplus$ & relativistic mass connection rule (ordinary addition)\\
$\uplus_0$ & rest mass connection rule (nontrivial, but equal to \\ & sum of relativistic masses in mass center frame)\\
$\mathrm{p}$ & the value of momentum\\
$\mathbf{p}$ &  momentum 3D vector\\
$\mathbf{p}_{sys}$ & momentum of the variable mass body system \\ & with additional low mass\\
$p^{\mu}$ & four-momentum (mass-momentum) 4D vector  \\ & in contravariant index form\\
$\hat{p}$ & four-momentum (mass-momentum) 4D vector \\ & in derivation form\\
$p^0$ & c-time ($ct$) component of four-momentum vector \\ & (equal to $Mc=E/c$)\\
$p^t$ & time component of four-momentum (equal to $M$)\\
$\vec{p}$ & spatial 3D part of four-momentum vector \\
& (equal to $\mathbf{p}$)\\
$p_{\mu}$ & four-momentum (energy-momentum) 4D covector \\ & in covariant index form\\
$\check{p}$ & four-momentum (energy-momentum) 4D covector \\ & in differential form\\
$p_0$ & c-time ($ct$) component of four-momentum covector\\ & (equal to $Mc=E/c$)\\
$p_t$ & time component of four-momentum covector \\ 
& (equal to $E=Mc^2$)\\
$\psi$ & factor-function for minuend velocity in $\ominus_{\perp}$\\
$\mathrm{q}$ & physical invariant scalar quantity in spacetime (4D)\\
$Q$ & physical semiscalar quantity in space (3D)\\
& correspond with spacetime invariant $\mathrm{q}$\\
$q^{\mu}$ & four-vector of physical  quantity in spacetime\\
$q^0$ & c-time ($ct$) component of four-vector $q^{\mu}$\\
$\vec{q}$ & spatial 3D part of four-vector $q^{\mu}$\\
$\mathbf{Q}$ & three dimensional equivalent of quantity $q^{\mu}$\\

$\rho$ & radius of curvature\\
$\mathbf{r}$ & position vector 3D\\
$\sigma$ & four-velocity of basic frame (implicit index)\\
$\sigma^{\mu}$ & four-velocity of basic frame (contravariant index)\\
$\sigma_{\nu}$ & four-velocity of basic frame (covariant index)\\
$\sigma{(\vec{0})}$ & four-velocity $(c,0,0,0)$ of the stationary system\\
$S_{\mathbf{v}}^{\gamma^2}$ & stretching (directional scaling) in the direction \\ & of velocity $\mathbf{v}$ and scale $\gamma^2$\\
$\tau$ & self time\\
$\tau_0$ & fixed self time for velocity of instantaneous \\ & and uniform co-moving system \\

$t$ & time\\
$t'$ & Lorentz transformation of $t$\\
$t_0$ & fixed time for velocity of uniform and instantaneous \\ & co-moving system (parameter, not variable)\\
$\hat{\mathbf{t}}$ & tangent versor of Frenet base\\
$\theta$ &  the value of the angle between force and velocity \\

$\upsilon$ & four-velocity of the second body (implicit index)\\
$\upsilon^{\mu}$ & four-velocity of the second body (contravariant index)\\
$\upsilon_{\nu}$ & four-velocity of the second body (covariant index)\\
$\vec{\upsilon}$ & spatial 3D part od four-velocity $\upsilon^{\mu}$\\
$\upsilon \cdot \nu$ & scalar product of 4D vectors equal \\ & $\upsilon_{\mu}\nu^{\mu}=\upsilon_0\nu^0+\upsilon_1\nu^1+\upsilon_2\nu^2+\upsilon_3\nu^3$ \\ & $\upsilon_{\mu}\nu^{\mu}=\upsilon^0\nu^0-\upsilon^1\nu^1-\upsilon^2\nu^2-\upsilon^3\nu^3$\\
$\mathrm{u}$ & the value of velocity $\mathbf{u}$\\
$\mathbf{u}$ & three dimensional velocity correspond to $\upsilon^{\mu}$\\
& (is not equal to $\vec{\upsilon}$)\\
$\mathbf{u}_0$ & relative velocity based on Einstein subtraction $\ominus_0$\\
$\mathbf{uv}$ & scalar product of 3D vectors\\
$U$ & capacitor voltage\\

$\nu$ & four-velocity of the first body (implicit index)\\
$\nu^{\mu}$ & four-velocity of the first body (contravariant index)\\
$\nu_{\nu}$ & four-velocity of the first body (covariant index)\\
$\vec{\nu}$ & spatial 3D part of four-velocity $\nu^{\mu}$\\
$\mathrm{v}$ & the value of velocity $\mathbf{v}$\\
$\mathrm{v}'$ & recompiled electron velocity value\\
$\mathbf{v}$ & three dimensional velocity correspond to $\nu^{\mu}$\\
& (is not equal to $\vec{\nu}$)\\

$\omega^{\mu}$ & Oziewicz binary 4D velocity of $\upsilon^{\mu}$ relative to $\nu^{\mu}$\\ 
$\omega^0$ & c-time ($ct$) component of $\omega^{\mu}$\\
$\vec{\omega}$ & spatial 3D part of $\omega^{\mu}$\\
$\omega^2$ & square of binary velocity equal to $\omega_{\mu}\omega^{\mu}=-\mathbf{u}^2_0$\\
$\omega^{\mu}_1$ & binary velocity of $\nu^{\mu}$ relative to $\sigma^{\mu}$\\ 
$\omega^{\mu}_2$ & binary velocity of $\upsilon^{\mu}$ relative to $\sigma^{\mu}$\\
$\omega(\tau_0, \tau)$ & binary four-velocity of the body in self time $\tau$ \\ & relative to the four-velocity of the body in time $\tau_0$\\

$\mathbf{w}$ & jet-type relative velocity based on $\ominus_{\perp}$ and three\\
& dimensional equivalent of $\omega^{\mu}$ (not equal to $\vec{\omega}$)\\
$\mathbf{w}_1$ & axial-type relative velocity based exactly on $\ominus_{\parallel}$\\
& or simpler based on $\ominus$ or $\ominus_1$ in Frenet base\\
$\mathbf{w}(t_0, t)$ & jet-type velocity of the body in time $t$ relative to \\ & the velocity of the body in time $t_0$\\

$\mathrm{W}$ & Searle electromagnetic energy of the moving charge\\
$W$ & Einstein's work of accelerating the body\\
$\mathbf{W}$ & Dragan ternary-type relative velocity based on $\ominus_{\wedge}$\\
& and 3D equivalent of $\xi^{\mu}$ (not equal to $\vec{\xi}$)\\
$\xi^{\mu}$ & Oziewicz ternary 4D velocity of $\upsilon^{\mu}$ relative to $\nu^{\mu}$\\ 
& in related to reference system $\sigma^{\mu}$\\
$\xi^0$ & c-time ($ct$) component of $\xi^{\mu}$\\
$\vec{\xi}$ & spatial 3D part of $\xi^{\mu}$\\
$(\xi^{\mu})_{\{\sigma\}}$ & ternary velocity $\xi^{\mu}$ in rest frame for $\sigma^{\mu}$\\ 
$\vec{\xi}_{\{\sigma\}}$ & vector $\vec{\xi}$ in rest frame for $\sigma^{\mu}$ (equal to $\mathbf{W}$)\\ 
$x$ & the first position coordinate\\
$\tilde{x}$ & Galilean transformation of $x$\\
$x'$ & Lorentz transformation of $x$\\
$x^{\mu}$ & four-position 4D vector\\
$\vec{x}$ & spatial 3D part od $x^{\mu}$ (equal to $\mathbf{r}$)\\
$\dot{x}^{\mu}$ & derivation of $x^{\mu}$ with respect to self time\\
& (equal to four-velocity $\nu^{\mu}$)\\
$x_1$ & distance of the hole from the radioactive source\\
$x_2$ & distance from the hole to the photographic plate\\
$y$ & the second position coordinate\\
$y'$ & Lorentz transformation of $y$ (equal $y$ here)\\
$y_0$ & coordinate of the electron on the photographic plate\\ &
(conditioned by electric field)\\
$z$ & the third position coordinate\\
$z'$ & Lorentz transformation of $z$ (equal $z$ here)\\
$z_0$ & coordinate of the electron on the photographic plate\\ &
(conditioned by magnetic field)\\
  \end{longtable}


\begin{thebibliography}{99}

\bibitem{Abraham 1} M. Abraham, ``Dynamik des Electrons (Dynamics of Electrons)", G\"ottinger Nachrichten,  pp. 20--41, January 11 (1902).

\bibitem{Abraham 2} M. Abraham, ``Prinzipien der Dynamik des Elektrons (Principles of the Dynamics of the Electrons)", Physikalische Zeitschrift 4 (1b), pp. 57--62 (1902).

\bibitem{Abraham 3} M. Abraham, ``Prinzipien der Dynamik des Elektrons (Principles of the Dynamics of the Electrons)", Annalen der Physik 10, pp. 105--179 (1903).

\bibitem{Ackeret} J. Ackeret, ``Zur Theorie der Raketen (On the theory of rockets)", Helvetica Physica Acta  19, pp. 103--112 (1946).

\bibitem{Adler} C. G. Adler, ``Does mass really depend on velocity, dad?", American Journal of Physics 55 (8), pp. 739--743 (1987).

\bibitem{Sudbury} B. Aharmim, S. N. Ahmed, A. E. Anthony, N. Barros, E.W. Beier, A. Bellerive, B. Beltran, M. Bergevin, S. D. Biller et al., ``Tests of Lorentz invariance at the Sudbury Neutrino Observatory", Physical Review D 98, 112013 (2018).

\bibitem{Bazanski} S. L. Ba\.za\'nski, ``Powstawanie i wczesny odbi\'or
szczeg\'olnej teorii wzgl\c{e}dno\'sci (Formation and early reception of special relativity)", Post\c{e}py Fizyki 56 (6), pp. 253--267 (2005).

\bibitem{Barrett} J.F. Barrett, ``Review of problems of dynamics in the hyperbolic theory of special relativity", PIRT Conf., Imperial Coll., London, Proceedings ISBN 1 873 694 07 5, PD Publications (Liverpool), pp. 17--30 (2002). 

\bibitem{Bowler} M. G. Bowler, ``Gravitation and Relativity", Pergamon, Elmsford, New York (1976).

\bibitem{Brillouin} L. Brillouin, ``Relativity Reexamined", Academic Press, New York, London (1970).

\bibitem{Bucherer 1904} A. H. Bucherer, ``Mathematische Einf\"uhrung in die Elektronentheorie (Mathematical introduction to the theory of electrons)", Verlag von B. G. Teubner, Leipzig (1904).

\bibitem{Bucherer} A. H. Bucherer, ``Messungen an Becquerelstrahlen. Die experimentelle Best\"atigung der Lorentz-Einsteinschen Theorie. (Measurements of Becquerel rays. The Experimental Confirmation of the Lorentz-Einstein Theory)", Physikalische Zeitschrift 9 (22), pp. 755--762 (1908).

\bibitem{Buquoy 1812} G. von Buquoy, ``Analytische Bestimmung des Gesetzes der Virtuellen Geschwindigkeiten in Mechanischer und Statischer Hinsicht (Analytical determination of the law of virtual velocities in mechanical and static terms)", Breitkopf und Hartel, Leipzig (1812).

\bibitem{Buquoy 1815} G. von Buquoy, ``Exposition d’un nouveau principe g\'en\'eral de dynamique, d’ont le principe des vitesses virtuelles n’est qu’un cas particulier (Exhibition of a new general principle of dynamics, have the principle of virtual velocities is only a particular case)", V Courcier, Paris (1815).

\bibitem{Cannoni} M. Cannoni, ``Lorentz invariant relative velocity and relativistic binary collisions", arXiv:1605.00569v2 (2016).

\bibitem{Criado} C. Criado, N. Alamo, ``From $E=mc^2$ to the Lorentz transformations via the law of addition of relativistic velocities", European Journal of Physics 26, pp. 611--616 (2005).

\bibitem{Cushing} J. T. Cushing, ``Electromagnetic mass, relativity, and the Kaufmann experiments", American Journal of Physics 49 (12), pp. 1133--1149 (1981).

\bibitem{Damour} T. Damour, ``Poincar\'e, the Dynamics of the
Electron, and Relativity", arXiv:1710.00706v1 (2017).

\bibitem{Dragan} A. Dragan, ``Niezwykle szczeg\'olna teoria wzgl\c{e}dno\'sci. Roz. 3. Obr\'ot Thomasa--Wignera  (Unusually special theory of relativity. Chap. 3. Thomas--Wigner rotation)", monograph -- lecture notes, www.researchgate.net/publication/265887295 (2012).

\bibitem{Dragan AJP} A. Dragan, T. Odrzyg\'o\'zd\'z, ``Half-page derivation of the Thomas precession", American Journal of Physics 81 (8), 631 (2013).

\bibitem{Einstein June} A. Einstein, ``Zur Elektrodynamik bewegter K\"orper (On the Electrodynamics of Moving Bodies)",
Annalen der Physik 17, pp. 891--921, June 30 (1905).


\bibitem{Einstein 1905} A. Einstein, ``Ist die Tr\"agheit eines K\"orpers von seinem Energieinhalt abh\"angig? (Does the Inertia of a Body Depend
Upon Its Energy-Content?)", Annalen der Physik 18, pp. 639--643, September 27 (1905).

\bibitem{Einstein 1906} A. Einstein, ``\"Uber eine Methode zur Bestimmung des Verh\"altnisses der transversalen und longitudinalen Masse des Elektrons (On a method for determination ratio of transverse and longitudal mass of electron)", Annalen der Physik 21 (13), pp. 583--586 (1906).

\bibitem{Einstein 1907} A. Einstein, ``\"Uber die vom Relativit\"atsprinzip geforderte Tr\"agheit der Energie (On the inertia of energy required by the relativity principle)", Annalen der Physik 23 (328/7), pp. 371--384 (1907).

\bibitem{Einstein 1935} A. Einstein, ``Elementary Derivation of the Equivalence of Mass and Energy", American Mathematical Society Bulletin 41, pp. 223--230 (1935).

\bibitem{Einstein 1946} A. Einstein, ``$E=mc^2$: The Most Urgent Problem of Our Time", Science Illustrated (The manuscript and reprints have survived) (1946).

\bibitem{Mexico} M. Fern\'andez-Guasti, ``Alternative realization for the composition of relativistic
velocities", Proc. of SPIE Vol. 8121, 812108 (2011).

\bibitem{Feynman} R. Feynman, R. B. Leighton, M. Sands, 
``The Feynman Lectures on Physics", Vol. 1, Sec. 15-1 ``The principle of relativity", Sec. 15-9 ``Equivalence of mass and energy", Sec. 16-4 ``Relativistic mass", Addison-Wesley (1963, 2005), Basic Books (2011).

\bibitem{Field} J. H. Field, `Einstein and Planck on mass-energy equivalence in 1905-06: a
modern perspective", European Journal of Physics 35 (5) (2014).

\bibitem{Field 2018} J. H. Field, `Feynman's dynamical route to special relativity via work-to-energy conversion and Newton's second law", Fundamental Journal of Modern Physics 11 (2), pp. 191--226 (2018).

\bibitem{Flores} F. Flores, ``Einstein's 1935 Derivation of $E=mc^2$", Studies in History and Philosophy of Modern Physics 29 (2), pp. 223--243 (1998).

\bibitem{Fock} V. Fock, ``The Theory of Space, Time and  Gravitation", 1st ed. 1959, 2nd rev. ed., translated by N. Kemmer, Pergamon Press (1964).

\bibitem{Franklin} J. Franklin, ``The lack of rotation in a moving right
angle lever", European Journal of Physics 29, pp. N55-–N58 (2008).

\bibitem{Gluza} J. Gluza, ``Relikt w fizyce -- poj\c{e}cie masy relatywistycznej (Relic in physics -- concept of relativistic mass)", Fizyka w Szkole 5, 269 (1992).

\bibitem{Hecht} E. Hecht, ``Einstein Never Approved of Relativistic Mass", The Physics Teacher 47, September (2009).

\bibitem{Hawking 1} S. Hawking, ``A Brief History of Time: From the Big Bang to Black Holes", Bantam Books (1988).

\bibitem{Hawking 2} S. Hawking, ``The Universe in a Nutshell", Bantam Spectra (2001).

\bibitem{ilusion} A. Jaffe, ``The illusion of time", Nature 556 (7701), pp. 304--305 (2018).

\bibitem{Janssen} M. Janssen, M. Mecklenburg, ``Electromagnetic Models of the Electron and the Transition from Classical to Relativistic Mechanics", text from conference: The Interaction between
Mathematics, Physics and Philosophy from 1850 to 1940, Carlsberg Academy, Copenhagen, September 26--28 (2002).

\bibitem{Johnson} W. Johnson, ``Contents and commentary on William Moore's A treatise on the motion of rockets and an essay on naval gunnery", International Journal of Impact Engineering 16 (3), pp. 499--521, June (1995).

\bibitem{Kaufmann 1901} W. Kaufmann, ``Die magnetische und elektrische Ablenkbarkeit der Becquerelstrahlen und die scheinbare Masse der Elektronen (The  magnetic  and  electrical
deflectability of Becquerel rays and the apparent mass of electrons)", G\"ottinger Nachrichten (2), pp. 143--155 (1901).  

\bibitem{Kaufmann 1902} W. Kaufmann, ``\"Uber die elektromagnetische Masse des Elektrons (On the Electromagnetic Mass of the Electrons)", G\"ottinger Nachrichten (5), pp. 291--296, July 26 (1902).

\bibitem{Kaufmann 1902 X} W. Kaufmann, ``Die elektromagnetische Masse des Elektrons (The Electromagnetic Mass of the Electrons)", Physikalische Zeitschrift 4 (1b), pp. 54--57, October 10 (1902).

\bibitem{Kaufmann 1903} W. Kaufmann, ``\"Uber die Elektromagnetische Masse der Elektronen (On the Electromagnetic Mass of Electron)", G\"ottinger Nachrichten (3), pp. 90--103 (1903).

\bibitem{Kaufmann 1905} W. Kaufmann, ``\"Uber die Konstitution des Elektrons (On the Constitution of the Electrons)", Sitzungsberichte der K\"oniglich Preussischen Akademie der Wissenschaften (45), pp. 949--956 (1905).

\bibitem{Kaufmann 1906} W. Kaufmann, ``\"Uber die Konstitution des Elektrons (On the Constitution of the Electron)", Annalen der Physik 19, pp. 487--553 (1906).

\bibitem{Koczan} G. Koczan, ``Wyprowadzanie promieniowania Hawkinga: Cz\c{e}\'s\'c II. Mechanika kwantowa oraz statystyczna stan\'ow fotonowych (Derivation of Hawking Radiation: Part II. Quantum and statistical mechanics of photon states)", Foton 141 Lato, pp. 4--32 (2018), English version: www.researchgate.net/publication/330369679 (2019).

\bibitem{Kostelecky} A. Kostelecky, N. Russell, ``Data Tables for Lorentz and CPT Violation", Review of Modern Physics 83, 11 (2011), https://arxiv.org/abs/0801.0287v13 (2020).

\bibitem{Kosyakov} B. P. Kosyakov, ``On inert properties of particles in classical theory", arXiv:hep-th/0208035v1 (2002).

\bibitem{Landau} L. D. Landau, E. M. Lifshitz, ``Teoriya polya (The Classical Theory of Fields)", Russian (1941, 1948), English: Addison-Wesley (1951), Pergamon Press (1959).

\bibitem{Langevin} P. Langevin, ``La physique des \'electrons (The physics of electrons)", Revue g\'en\'erale des sciences pures et appliqu\'ees 16 (6), pp. 257--276, March 30 (1905).

\bibitem{Lammerzahl} C. L\"ammerzahl, ``Special Relativity and Lorentz Invariance", Annalen der Physik 14 (1-3), pp. 71--102 (2005).

\bibitem{Lewis 1908} G. N. Lewis, ``A revision of the Fundamental Laws of Matter and Energy", Philosophical Magazine S. 6 Vol. 16 No. 95, pp. 705--717 (1908).

\bibitem{Lewis} G. N. Lewis, R. C. Tolman, ``The Principle of Relativity, and Non-Newtonian Mechanics", Proceedings of the American Academy of Arts and Sciences 44 (25), pp. 709--726 (1909).

\bibitem{Lorentz 1899} H. A. Lorentz, ``Simplified Theory of Electrical and Optical Phenomena in Moving Systems", Proceedings of the Royal Netherlands Academy of Arts and Sciences 1, pp. 427--442 (1899).

\bibitem{Lorentz 1904} H. A. Lorentz, ``Electromagnetic phenomena in a system moving with any velocity smaller than that of light", Proceedings of the Royal Netherlands Academy of Arts and Sciences 6, pp. 809-831 (1904).

\bibitem{Lorentz 1914} H. A. Lorentz, ``Das Relativit\"atsprinzip. Drei Vorlesungen gehalten in Teylers Stiftung zu Haarlem. (The Relativity Principle. Three lectures held at the Teylers Foundation in Haarlem.)", B. G. Teubner, Leipzig and Berlin (1914).

\bibitem{Mach} E. Mach, ``Die Mechanik in ihrer Entwicklung - Historisch kritisch dargestellt (The Science of Mechanics - A Critical and Historical Account of Its Development)", first German (1883), first English (1893), fourth English: T. J. McCormack, The Open Court Publishing CO. (1919). 

\bibitem{Mansouri} R. Mansouri, R. U. Sexl, ``A Test Theory of Special Relativity: I. Simultaneity
and Clock Synchronization", General Relativity and Gravitation 8 (7), pp. 497--513 (1977).

\bibitem{Meissner} K. A. Meissner, ``$E=mc^2$", multimedia lecture in National Centre for Nuclear Research May 23 (2011) and Faculty of Physics University of Warsaw (2014), 
https://www.youtube.com/watch?v=\_VDAy9zyvZ0 (2016).

\bibitem{Mieszczerski 1897} I. V. Meshchersky, ``Dinamika tochki peremnnoy massy (Dynamics of a point of variable mass)", Akademia Nauk, Peterburskij Universitet, St Petersburg (1897).

\bibitem{Mieszczerski 1904} I. V. Meshchersky, ``Uravneniya dvizheniya tochki peremennoy massy v obshchem sluchaye (Equations of motion of a variable mass point in the general case)", St. Petersburg Polytechnic University News 1, pp. 77--118 (1904).

\bibitem{Minkowski} H. Minkowski, ``Die Grundgleichungen für die elektromagnetischen Vorg\"ange in bewegten K\"orpern (The Fundamental Equations for Electromagnetic Processes in Moving Bodies)", Nachrichten von der Gesellschaft der Wissenschaften zu G\"ottingen, Mathematisch-Physikalische Klasse, pp. 53--111, Berlin (1908). 

\bibitem{Moore} W. Moore, ``A Treatise on the Motion of Rockets and An Essay on Naval Gunnery", The Military Academy at Woolwich, G. \& S. Robinson, London (1813).

\bibitem{Morozov} V. B. Morozov, ``On the question of the electromagnetic momentum of a charged body", Physics--Uspekhi 54, pp. 371--374 (2011).

\bibitem{First test} S. E. M\"uller, E. A. Dijck, H. Bekker, J. E. van den Berg, O. B\"oll, S. Hoekstra, K. Jungmann, C. Meinema, J. P. Noordmans et al., ``First Test of Lorentz Invariant in the Weak Decay of Polarized Nuclei", Physical Review D 88, 071901 (2013).

\bibitem{Spin 2} A. Naruko, R. Kimura, D. Yamauchi, ``On Lorentz-invariant spin-2 theories", Physical Review D 99, 084018 (2019).

\bibitem{Newton} I. Newton, ``Philosophiae Naturalis Principia Mathematica (Mathematical Principles of Natural Philosophy) (Matematyczne zasady filozofii przyrody)", London (1687), A. Motte (1729), J. Wawrzycki, Copernicus Center Press, Cracow (2011).

\bibitem{Nowik} A. Nowik, ``Prawda jest jedna, a g\l{}upstw tysi\c{a}ce -- uwagi do dyskusji o masie relatywistycznej 
(The truth is one, and the lies are thousands -- remarks on the discussion about the relativistic mass)", Fizyka w Szkole z Astronomi\c{a} 4, pp. 33--34 and 37 (2016). 

\bibitem{Oas} G. Oas, ``On the abuse and use of relativistic mass", arXiv:physics/0504110v2 (2005).

\bibitem{Okun} L. B. Okun, ``The Concept of Mass", Physics Today 42 (6), pp. 31--36 (1989).

\bibitem{Okun 2} L. B. Okun, ``Energy and Mass in Relativity Theory", World Scientific (2009).

\bibitem{Okun 3} L. B. Okun, ``Mass versus relativistic and rest masses", American Journal of Physics 77 (5), pp. 430--431 (2009).

\bibitem{Osiak} Z. Osiak, ``Energy in Special Relativity", Theoretical Physics 4(1), pp. 22--25,
Isaac Scientific Publishing (2019), www.researchgate.net
/publication/322626086 (2018), first preprint 1512.0449 (2015).

\bibitem{binary} Z. Oziewicz, ``How do you add relative velocities?", Group Theoretical Methods in Physics, Conference Series Number 185, pp. 439--444,  CRC Press (2004).

\bibitem{Oziewicz} Z. Oziewicz, ``Ternary relative velocity", arXiv:1104.0682v1 (2011).

\bibitem{groupoid} Z. Oziewicz, ``Relativity groupoid, instead of relativity group", International Journal of Geometric Methods in Modern Physics 04 (05), pp. 739--749 (2007).

\bibitem{without Lorentz} Z. Oziewicz, ``Relativity without Lorentz group", monograph, www.academia.edu/17230451 (2004, 2007).

\bibitem{Penrose} R. Penrose, ``The Road to Reality", Chap. 18.6--18.7, Jonathan Cape, London (2004--2007). 

\bibitem{Pesce} C. P. Pesce, L. Casetta, ``Variable mass systems dynamics in engineering mechanics education", Proceedings of COBEM, 19th International Congress of Mechanical Engineering,  Brasilia (2007).

\bibitem{Planck} M. Planck, ``Das Prinzip der Relativit\"at und die
Grundgleichungen der Mechanik (The Principle of Relativity and the Fundamental Equations of Mechanics)", Verh. D. Phys. Ges. 8, pp. 136--141, March 23 (1906).

\bibitem{Planck mass} M. Planck, ``Die Kaufmannschen Messungen der Ablenkbarkeit der $\beta$-Strahlen in ihrer Bedeutung f\"ur die Dynamik der Elektronen (The Measurements of Kaufmann on the Deflectability of $\beta$-Rays in their Importance for the Dynamics of the Electrons)", Physikalische Zeitschrift 7, pp. 753--761, September 19 (1906).

\bibitem{Poincare} H. Poincar\'e, ``Sur la dynamique de l’\'electron (On the Dynamics of the Electron)", Comptes rendus hebdomadaires des s\'eances de l'Acad\'emie des sciences 140, pp. 1504--1508, June 5 (1905).

\bibitem{Poisson} S. D. Poisson, ``Sur le mouvement d’un systeme de corps, en supposant les masses variables (On the motion of a body system, assuming variable masses)", Bull. Sci. Soc. Philomat. 60-2, Paris (1819).


\bibitem{Rebilas} K. R\c{e}bilas, ``Derivation of the relativistic
momentum and relativistic equation of motion from Newton's second law and Minkowskian space-time
geometry", Apeiron 15 (3), July (2008).

\bibitem{Rebilas 2} K. R\c{e}bilas, ``Comment on ‘Elementary analysis of
the special relativistic combination of velocities, Wigner rotation and Thomas precession’ ", European Journal of Physics 34, pp. L55--L61 (2013).

\bibitem{Rebilas 3} K. R\c{e}bilas, ``Lorentz-invariant three-vectors and alternative formulation of relativistic dynamics", American Journal of Physics 78, 294 (2010).

\bibitem{Rindler 1} W. Rindler, ``Introduction to Special Relativity", Oxford University Press (1982).

\bibitem{Rindler 2} W. Rindler, ``Relativity: Special, General and Cosmological", Oxford University Press (2001, 2006).

\bibitem{putting rest} W. Rindler, M. A. Vandyck, P. Murugesan, S. Ruschin, C. Sauter, L. B. Okun, ``Putting to Rest Mass Misconceptions", Physics Today, Letters 43 (5), pp. 13--14, 115, 117, May (1990).

\bibitem{Roche} J. Roche, ``What is mass?", European Journal of Physics 26, pp. 1--18 (2005).

\bibitem{Rovelli} C. Rovelli, ``The Order of Time", Penguin Books (2018).

\bibitem{Sandin} T. R. Sandin, ``In defense of relativistic mass", American Journal of Physics 59 (11),  pp. 1032--1036 (1991).

\bibitem{Savrov} L. A. Savrov, ``Gravitational Shielding and the Equivalence Principle", Gravitation and Cosmology 18 (4), pp. 270--278 (2012).


\bibitem{Searle} G. F. C. Searle, ``On the Steady Motion of an Electrified Ellipsoid", Philosophical Magazine 5, 44(269), pp. 329--341 (1897).

\bibitem{Sturm} S. Sturm, F. K\"ohler, J. Zatorski, A. Wagner, Z. Harman, G. Werth, W. Quint, C. H. Keitel, K. Blaum, ``High-precision measurement of the atomic mass of the electron", Nature 506, pp. 467--470 (2014).

\bibitem{reference system} K. Szostek, R. Szostek, ``The derivation of the general form of kinematics with the universal reference system", Results in Physics 8, pp. 429--437 (2018).

\bibitem{Taylor} E. F. Taylor, J. A. Wheeler, ``Spacetime physics - introduction to special relativity", W.H. Freeman and Company (1966, 1992).

\bibitem{Tangherlini} F. R. Tangherlini, ``The velocity of light in uniformly moving frame, A dissertation", Stanford University  (1958), The Abraham Zelmanov Journal Vol. 2, pp. 44--110 (2009).

\bibitem{Thomas} L. H. Thomas, ``Motion of the spinning electron", Nature 117, 514 (1926).

\bibitem{Ciolkowski} K. Tsiolkovsky, ``Issledovaniye mirovykh prostranstv reaktivnymi priborami (Exploration of world spaces by reactive devices)", A Scientific-Philosophical and Literary Journal (in Russian) No. 5, St Petersburg May (1903).

\bibitem{Tolman} R. Tolman, ``Non-Newtonian Mechanics. The Mass of a Moving Body", Philosophical Magazine 23(135), pp. 375--380 (1912).

\bibitem{Ugarov} V. A. Ugarov, ``Special Theory of Relativity", Russian (1969), English Y. Atanov Mir (1979).

\bibitem{Ungar 1997} A. A. Ungar, ``Thomas precession: its underlying gyrogroup axioms and their use in hyperbolic geometry and relativistic physics", Foundations of Physics 27, pp. 881-–951 (1997).

\bibitem{Ungar 2007} A. A. Ungar, ``Gyrogroups, the Grouplike Loops in the Service of Hyperbolic Geometry and Einstein's Special Theory of Relativity", Quasigroups and Related Systems 15, pp. 141--168 (2007).

\bibitem{Wolny}  J. Wolny, R. Strza\l{}ka, ``Momentum in the Dynamics of Variable-Mass Systems:
Classical and Relativistic Case", Acta Physica Polonica A 135(3), pp. 475--479 (2019).

\bibitem{Wroblewski historia} A. K. Wr\'oblewski, ``Historia fizyki -- od czas\'ow najdawniejszych do wsp\'o\l{}czesno\'sci (History of physics -- from the earliest times to the present day)", PWN (2007).

\bibitem{Wroblewski} A. K. Wr\'oblewski, J. A. Zakrzewski, ``Wst\c{e}p do fizyki (Introduction to physics)" Vol. 1, PWN (1976, 1984).

\bibitem{Zatorski} J. Zatorski, B. Sikora, S. G. Karshenboim, S. Sturm,
F. K\"ohler-Langes, K. Blaum, C. H. Keitel, Z. Harman, ``Extraction of the electron mass from $g$ factor measurements on light hydrogenlike ions", Physical Review A 96, 012502 (2017).

\bibitem{Gravitational mass} M. Zych, \L{}. Rudnicki, I. Pikovski, ``Gravitational mass of composite systems", Physical Review D 99, 104029 (2019).


\end{thebibliography}
\end{document}